\documentclass[epj,nopacs,fleqn]{svjour}

\usepackage{listings}
\newcommand{\n}{\nonumber}

\newcommand{\eps}{\epsilon}
\newcommand{\del}{\partial}

\newcommand{\sgn}{{\rm sgn}}
\newcommand{\vect}{\overrightarrow}

\newcommand{\qvec}{\overrightarrow q}

\usepackage{amsmath}
\usepackage{kantlipsum}
\allowdisplaybreaks

\usepackage{graphics}
\usepackage{amsmath,amssymb,bbm,bm}
\usepackage{slashed}
\usepackage{hyperref}
\usepackage{color}
\usepackage{tikz-feynhand}
\usepackage{cite}
\usepackage{multirow}
\usepackage{comment}
\usepackage[super]{nth}

\begin{document}

\title{
Helicity amplitudes in light-cone and Feynman-diagram gauges}

\author{
	Junmou Chen\inst{1,}\thanks{chenjm@jnu.edu.cn}
	\and
	Kaoru Hagiwara\inst{2,3}\thanks{kaoru.hagiwara@kek.jp}
	\and
	Junichi Kanzaki\inst{4,}\thanks{junichi.kanzaki@ipmu.jp}
	\and
	Kentarou Mawatari\inst{5,}\thanks{mawatari@iwate-u.ac.jp}
	\and
	Ya-Juan Zheng\inst{5,6,}\thanks{yjzheng@iwate-u.ac.jp}	
}                     

\institute{
	Department of Physics, Jinan University, Guangzhou, Guangdong Province, 510632, China
	\and 
	Center for Theory and Computation, National Tsing Hua University, Hsinchu, Taiwan 300
	\and
	KEK Theory Center and Sokendai, Tsukuba, Ibaraki 305-0801, Japan
	\and
	Kavli IPMU (WPI), UTIAS, The University of Tokyo, Kashiwa, Chiba 277-8583, Japan
	\and
	Faculty of Education, Iwate University, Morioka, Iwate 020-8550, Japan
	\and
	Department of Physics and Astronomy, University of Kansas, Lawrence, KS 66045, U.S.A.
}

\date{}

\abstract{
Recently proposed Feynman-diagram (FD) gauge propagator for massless and massive gauge bosons is
 obtained from a light-cone (LC) gauge propagator, by choosing the gauge vector along the opposite direction of the gauge boson three-momentum.
We implement a general LC gauge propagator for all the gauge bosons of the Standard Model (SM) in the HELicity Amplitude Subroutines ({\tt HELAS}) codes, such that all the SM helicity amplitudes can be evaluated at the tree level in the LC gauge by using {\tt MadGraph}.
We confirm that our numerical codes produce physical helicity amplitudes which are consistent among all gauge choices. 
We then study interference patterns among Feynman amplitudes, for a few $2\to3$ scattering processes in QED and QCD, and the process $\gamma\gamma\to W^+W^-$ followed by the $W^\pm$ decays.
We find that in a generic LC gauge, where all the gauge boson propagators share a common gauge vector, we cannot remove the off-shell current components which grow with their energy systematically from all the Feynman amplitudes in $2\to3$ processes.
On the other hand, the $5\times5$ LC gauge propagator for the weak bosons removes components which grow with energy due to the longitudinal polarization mode of the external bi-fermion currents, and hence can give $2\to2$ weak boson scattering amplitudes which are free from subtle cancellation at high energies.
The particular choice of the FD gauge vector has advantages over generic LC gauge, not only because all the terms which grow with energy of off-shell and on-shell currents are removed systematically from all the diagrams, but also because no artificial gauge vector direction dependence of individual amplitudes appears. 
} 

\authorrunning{CHKMZ}
\maketitle

\vspace*{-12cm}
\vspace*{10cm}

\vskip -12cm
\noindent KEK-TH-2471, IPMU22-0055\qquad  
\vspace*{12cm}

\section{Introduction}\label{sec:intro}

It has been shown recently~\cite{Hagiwara:2020tbx,Chen:2022gxv} that the tree-level scattering amplitudes of the Standard Model (SM) can
be expressed in a form which is free from subtle
gauge theory cancellation among interfering Feynman
diagrams.  
The specific form of the gauge boson propagators
which appear in the amplitudes are called
`Feynman Diagram (FD)' gauge\!\cite{Chen:2022gxv}, since the
helicity amplitudes corresponding to individual
Feynman diagram are expressed as products of
the invariant Feynman's propagator factors,
for the connecting lines, and the `splitting
amplitudes'\!\cite{Hagiwara:2009wt} at each vertex\footnotemark.
Splitting amplitudes in the Electroweak (EW)
theory for the weak bosons and the Higgs boson
are found e.g. in refs.\!\cite{Chen:2016wkt,Cuomo:2019siu}, where the FD
gauge propagators for the weak bosons are
obtained~\cite{Wulzer:2013mza} along studies of the Goldstone boson
equivalence theorem\!\cite{Chanowitz:1985hj,Veltman:1989ud}.  

In refs.\!\cite{Hagiwara:2020tbx,Chen:2022gxv}, the FD gauge propagators are
obtained from the scattering amplitudes
expressed in a covariant gauge, Feynman
or Landau gauge in QED and QCD~\cite{Hagiwara:2020tbx}, or
the Unitary gauge in the EW theory~\cite{Chen:2022gxv}.
%
\footnotetext{Because of this property, the authors of ref.\!\cite{Hagiwara:2020tbx}
called their photon and the gluon propagators
as `Parton Shower (PS) gauge'.
Since the term `PS gauge' had been used in the
parton shower studies\!\cite{Nagy:2007ty,Nagy:2014mqa}, we adopt the
naming `FD gauge', following ref.\!\cite{Chen:2022gxv}}
%
In case of the Landau gauge, the polarization
tensor of the gauge boson propagator
\begin{eqnarray}
 P_{j=1}^{\mu\nu} = -g^{\mu\nu}
+ \frac{q^\mu q^\nu}{q^2} 
\end{eqnarray}
can be expressed as a summation over the
polarization vectors of definite helicities
\begin{eqnarray}
P_{j=1}^{\mu\nu}
 =
 \sum_{h=\pm 1} \eps^\mu(q,h) \eps^\nu(q,h)^*
 +
 \sgn{(q^2)} \eps^\mu(q,0) \eps^\nu(q,0).
 \nonumber\\
 \label{eq:polsum}
\end{eqnarray}
 The FD gauge propagator is obtained by noting
 that the longitudinal polarization vector
 can be expressed as a sum of the scalar
 term and the difference
 \begin{eqnarray}
 \eps^\mu(q,0) = 
\frac{q^\mu}{Q}
 +
  \tilde\eps^\mu(q,0) ,
 \label{eq:pol_long}
 \end{eqnarray}
 where $Q=\sqrt{|q^2|}$ is the virtuality of
 the photon or gluon.  
 Inserting eq.\eqref{eq:pol_long} into eq.\eqref{eq:polsum}, and dropping
 all terms proportional to $q^\mu$ or $q^\nu$,
 by applying the BRST identities~\cite{Becchi:1975nq,Tyutin:1975qk} for the
 sub-amplitudes with only one off-shell gauge
 boson, we arrive at the polarization tensor
 of the FD gauge:
\begin{eqnarray}
P_{\rm FD}^{\mu\nu}
 =
 \sum_{h=\pm 1} \eps^\mu(q,h) \eps^\nu(q,h)^*
 +
 \sgn{(q^2)} \tilde\eps^\mu(q,0) \tilde\eps^\nu(q,0).
 \nonumber\\
 \label{eq:PFDmunu}
\end{eqnarray}

 In case of the EW theory, we start from the amplitudes
 in the Unitary gauge of the weak bosons, whose
 polarization tensor
\begin{eqnarray}
P_{\rm U}^{\mu\nu}(V) = -g^{\mu\nu}
 + \frac{q^\mu q^\nu}{m_V^2}
\end{eqnarray}
 for $V=W^\pm$ or $Z$ can be expressed as\!\cite{Chen:2022gxv}
 \begin{eqnarray}
 P_{\rm U}^{\mu\nu}(V)
 &=&
 \sum_{h=\pm 1} \eps^\mu(q,h) \eps^\nu(q,h)^\ast
+
 \sgn{(q^2)} \tilde\eps^\mu(q,0) \tilde\eps^\nu(q,0)
  \nonumber\\
 &+&
 \tilde\eps^\mu(q,0) \frac{q^\nu}{Q}
 +
 \frac{q^\mu}{Q} \tilde\eps^\nu(q,0)
 +
 \frac{q^\mu q^\nu}{m_V^2}\,.
 \end{eqnarray}
 Now, the BRST identities\!\cite{Becchi:1975nq,Tyutin:1975qk} for the sub-amplitudes
 give the corresponding Goldstone boson amplitudes,
 arriving at the $5\times 5$ representation of the
 FD gauge propagator
\begin{eqnarray}
P_{\rm FD}^{MN}(V) = 
  \begin{pmatrix}
	  {P}_{\rm FD}^{\mu\nu} & \displaystyle{-i{\tilde{\eps}}^\mu(q,0)\frac{m_V}{Q} } \\ 
	 \displaystyle{ i\tilde\eps^\nu(q,0)\frac{m_V}{Q} } &1
  \end{pmatrix}\!, 
  \label{eq:PFDMN}
\end{eqnarray}
 where the first 4 components of $M$ and $N=0,1,2,3$,
 denoted by $\mu$ and $\nu$, respectively,
 account for the four-vector components
 of the weak bosons, whereas their fifth
 components, $M,N=4$, account for the
 associated Goldstone boson.
 \footnote{The BRST identities between the matrix elements
 of the operator $\del^\mu V_\mu(x)$ and those
 of the Goldstone boson operator $\pi(x)$ at the tree level
 are $\xi$ independent~\cite{Gounaris:1986cr} in the covariant $R_\xi$
 gauge~\cite{Fujikawa:1972fe}.
 We can take the $\xi \to \infty$ limit smoothly
 to obtain the Unitary gauge expression, since
 all the Goldstone boson couplings are $\xi$
 independent.
}

In refs.\!\cite{Hagiwara:2020tbx} and\,\cite{Chen:2022gxv}, the above FD gauge propagators
 for the SM gauge bosons, eq.\,\eqref{eq:PFDmunu} for the photon and gluons and eq.\,\eqref{eq:PFDMN} for the $W^\pm$ and $Z$ bosons, have been implemented into
 the numerical helicity amplitude calculation codes
 {\tt HELAS}\!\cite{Hagiwara:1990dw,Murayama:1992gi}, and a few modifications were applied to
 the Feynman rules for some EW vertices~\cite{Chen:2022gxv}
 such that we can obtain helicity amplitudes of an
 arbitrary SM processes by using {\tt MadGraph5\_aMC@NLO}\!\cite{Stelzer:1994ta,Alwall:2011uj,Alwall:2014hca}
 at the tree level.
 The results presented in refs.\!\cite{Hagiwara:2020tbx,Chen:2022gxv} are encouraging,
 since not only all the numerical problems associated
 with so-called `gauge cancellation' among interfering
 Feynman amplitudes are absent, but also the helicity
 amplitudes for individual Feynman diagram seem to have
 definite physical interpretation, as products of all
 the invariant Feynman propagator factors for the
 connecting lines and the splitting amplitudes\!\cite{Hagiwara:2009wt,Chen:2016wkt,Cuomo:2019siu} 
 at each vertex.
  The latter property may allow us to study physics of
 interference among interfering Feynman amplitudes
 by using numerical codes.
 On the other hand, it has not been made clear how
 the FD gauge amplitudes are automatically generated
 for quantum field theory (QFT) models beyond the SM.
In addition, it was not clear if the FD gauge
 propagators can be used in loop calculations,
 because all what was done in refs.\!\cite{Hagiwara:2020tbx,Chen:2022gxv} is to
 replace the known covariant gauge propagators
 by the FD gauge ones by making use of the BRST
 identities for the two sub-amplitudes which are
 connected by a gauge boson line.  
  Because the property that cutting of an internal
 line gives two sub-amplitudes, each reduces to 
 scattering amplitudes when the cutted line is
 set on-shell, is a property unique to the tree-level
 amplitudes, the heuristic derivation given in
 refs.\!\cite{Hagiwara:2020tbx,Chen:2022gxv} cannot be applied to loop amplitudes.
 An alternative approach within the covariant quantization scheme of the electroweak theory has been pursued by the authors of refs.\cite{Cuomo:2019siu,Wulzer:2013mza}.  

 In this paper, we present an alternative derivation of the FD gauge propagators, in which they are obtained directly as
 Green's functions of the equation of motion of
 the free weak bosons quantized in the Light Cone
 (LC) gauge. 
By choosing the LC gauge vector along the opposite of
 the three-momentum direction of the vector boson\!\cite{Hagiwara:2020tbx,Chen:2022gxv}
 \begin{align}
 n_{\rm FD}^\mu = 
 \begin{pmatrix}
 \sgn{(q^0)}, &-\qvec/\left|\qvec\right|
 \end{pmatrix},
 \end{align}
 we obtain the FD gauge propagators.
 In this derivation, it is clear that the Goldstone
 boson degrees of freedom should be the \nth{5} component
 of the physical weak boson wave functions, and hence no 
 unphysical particles remain in the spectrum,
 at least in the tree level.
 The Feynman rules are then obtained straightforwardly in an arbitrary QFT models.

 The paper is organized as follows.
 In section~\ref{sec:deriv}, we show how the FD gauge propagator
 for the weak bosons can be obtained directly from
 their free equation of motion.
 In section~\ref{sec:results}, we show sample results for a few
 $2 \to 3$ processes in QED and QCD, comparing Feynman amplitudes in the 
 FD gauge against those in a few generic choices of LC gauges
 and the Feynman gauge.
 In section~\ref{sec:EW}, we show our findings for the amplitudes
 of the process,
 $\gamma\gamma\to W^+ W^-$ followed by $W^\pm$ decays,
 where we compare the FD gauge amplitudes against
 those of a few generic LC gauges and the Unitary gauge.
 Section~\ref{sec:summary} summarizes our findings, and outlines how
 automatic computation in the FD gauge is possible
 for BSM models, including SM Effective Field Theory (SMEFT).

\section{Derivation}\label{sec:deriv}

The free Lagrangian for the $Z$ boson and the corresponding Goldstone boson ($\pi$) 
in the LC gauge is given by
\begin{align}
  {\cal L}^{(0)}&=
  -\frac{1}{4}(\del^\mu Z^\nu-\del^\nu Z^\mu)^2 +\frac{1}{2}m^2Z^\mu Z_\mu
   \n\\
  &\quad +\frac{1}{2}(\del^\mu\pi)^2
  +mZ^\mu\del_\mu\pi 
  -\frac{1}{2\xi}(n^\mu Z_\mu)^2, 
  \label{eq:L0}
\end{align} 
where $m$ is the $Z$ boson mass, and $\xi$ and $n^\mu$ are the gauge-fixing parameters, satisfying $n\cdot n=0$.
The above Lagrangian can be expressed as
\begin{eqnarray}
  &&{\cal L} ^{(0)}=\frac{1}{2}\times
    \n\\	
 && \begin{pmatrix}
	  Z^\mu~ & \pi
  \end{pmatrix}
  \begin{pmatrix}
	  (\del^2+m^2)g_{\mu\nu}-\del_\mu\del_\nu-\displaystyle{\frac{n_\mu n_\nu }{\xi}}~~
    & m\del_\mu \\
	    -m\del_\nu & -\del^2 \\
  \end{pmatrix}\!\!
  \begin{pmatrix}
	  Z^\nu \\ \pi
  \end{pmatrix}\!,  
  \nonumber\\
\label{L_Z}
\end{eqnarray}
with the 5-component fields ($Z^\mu,\pi$), where the derivatives operate to the right-hand side.

The equation of motion (EOM) of the $(Z^\nu, \pi)$ system is then
\begin{eqnarray}
  &&\begin{pmatrix}
	  (\del^2+m^2)g^{\mu}_{\ \nu}-\del^\mu\del_\nu-\displaystyle{\frac{n^\mu n_\nu }{\xi}}~~
    & m\del^\mu \\
	    -m\del_\nu  & -\del^2 \\
  \end{pmatrix}
  \begin{pmatrix}
	  Z^\nu(x) \\ \pi(x)
  \end{pmatrix}
   \nonumber\\
 &&=0.  
\label{eom}
\end{eqnarray}
In the momentum space, the EOM reads
\begin{eqnarray}
 && \begin{pmatrix}
	  (-q^2+m^2)g^{\mu}_{\ \nu}+q^\mu q_\nu-\displaystyle{\frac{n^\mu n_\nu }{\xi}}~~
    & -i m q^\mu \\
	    i m q_\nu  & q^2 \\
  \end{pmatrix}
  \begin{pmatrix}
	  \tilde Z^\nu(q)  \\ \tilde\pi(q)
  \end{pmatrix}
    \n\\
  &&=0,  
\label{eom_mom}
\end{eqnarray}
where $\tilde Z^\nu(q)$ and $\tilde\pi(q)$ are the Fourier transform of 
$Z^\nu(x)$ and $\pi(x)$ fields.
We can express the EOM~\eqref{eom_mom} as
\begin{align}
 O^M_{\ N}\tilde Z^N(q)=0
\end{align}
with $M,N=0$ to 4, where 0 to 3 are $\mu,\nu$, and 
\begin{align}
  \tilde Z^4(q)=\tilde Z_4(q)=\tilde\pi(q).
\end{align}
The $5\times5$ matrix $O$ is
\begin{subequations}
\begin{align}
 O^\mu_{\ \nu}&=(-q^2+m^2)g^{\mu}_{\ \nu}+q^\mu q_\nu-\frac{n^\mu n_\nu }{\xi}, \\
 O^\mu_{\ 4}  &=-i m q^\mu, \\
 O^4_{\ \nu}  &=i m q_\nu, \\
 O^4_{\ 4}    &=q^2.
\end{align}
\end{subequations}
It is then straightforward to show the following identity:
\begin{align}
 P^R_{\ M}O^M_{\ N}\tilde Z^N(q)=(q^2-m^2)\delta^R_{\ N}\tilde Z^N(q)=0
\label{inverseeom} 
\end{align}
with the $5\times5$ matrix
\begin{subequations}
\begin{align}
 P^\rho_{\ \mu}&=-g^{\rho}_{\ \mu}+\frac{n^\rho q_\mu+q^\rho n_\mu }{n\cdot q}
                 +\xi(q^2-m^2) \frac{q^\rho q_\mu}{(n\cdot q)^2}, \\
 P^\rho_{\ 4}  &= i\frac{mn^\rho}{n\cdot q}
                 +i\xi(q^2-m^2) \frac{mq^\rho}{(n\cdot q)^2}, \\
 P^4_{\ \mu}   &=-i\frac{mn_\mu}{n\cdot q}
                 -i\xi(q^2-m^2) \frac{mq_\mu}{(n\cdot q)^2}, \\
 P^4_{\ 4}     &=1+\xi(q^2-m^2) \frac{m^2}{(n\cdot q)^2}.
\end{align}
\end{subequations}
The identity\,\eqref{inverseeom} tells that all 5-components 
of $\tilde Z^N(q)$ have a common Lorentz invariant mass,
satisfying the equation of motion~\eqref{eom_mom} at
an arbitrary Lorentz frame.
The LC gauge propagator of the weak boson is then obtained
directly as the Green's function of the EOM: 
\begin{align}
 G^R_{\ M} =\frac{P^R_{\ M}}{q^2-m^2+i\varepsilon}.
\label{gfunc} 
\end{align}

It should be noted that the above $5\times 5$ representation
of the weak boson propagators were obtained in 1987 by
Kunszt and Soper\!\cite{Kunszt:1987tk} in genral axial gauge.
They solve the equation of motion under the constraint,
\begin{eqnarray}
n^\mu V_\mu(x) = 0
\end{eqnarray}
for a generic constant four vector $n^\mu$ (axial gauge).
Our LC gauge Green's function~\eqref{gfunc} agrees with
the one reported in ref.\!\cite{Kunszt:1987tk} 
exactly when we set $\xi=n^2=0$. 

In the following, we set $\xi=0$ and keep only the light-cone vector $n^\mu$
\begin{align}
  n^\mu=(1,\vect{n}),\quad |\vect{n}|=1,
\end{align}
as the gauge parameter.
The `polarization' tensor in the LC gauge is then
\begin{align}
  {P}_{\rm LC}{}^M_{\ N}=
  \begin{pmatrix}
          -g^{\mu}_{\ \nu}+\dfrac{n^\mu q_\nu+q^\mu n_\nu }{n\cdot q} \ ~~
		& i m \dfrac{n^\mu}{n\cdot q} \\
          -i m \dfrac{n_\nu}{n\cdot q} & 1 \\
  \end{pmatrix},
  \label{pol5}
\end{align}
and the LC gauge propagator is 
\begin{align}
 G_{\rm LC}{}^M_{\ N} =\frac{P_{\rm LC}{}^M_{\ N}}{q^2-m^2+i\varepsilon}\,.
\label{gfunc_LC} 
\end{align}

 The FD gauge propagator is then obtained by setting 
\begin{align}
  n^\mu
  \equiv
  n^\mu_{\rm FD}
  =
  \begin{pmatrix}
   \sgn(q^0),&-\vect{q}/|\vect{q}|  
   \end{pmatrix}
   \label{eq:FDprop}
\end{align}
as defined in refs.\!\cite{Hagiwara:2020tbx,Chen:2022gxv}.
By comparing the LC gauge propagator factor\,(\ref{pol5}) with the corresponding FD gauge factor eq.\,\eqref{eq:PFDMN}, we find 
\begin{eqnarray}
\tilde{\epsilon}^\mu(q,0)=-{\rm sgn}(q^2)\,\frac{Q\,n^\mu_{\rm FD}}{n_{\rm FD}\cdot q},
\end{eqnarray}
as shown in ref.\!\cite{Chen:2022gxv}.
It should be noted here that, although the FD gauge propagators can be obtained from the LC gauge propagator~\eqref{gfunc_LC} simply by choosing the LC gauge vector as\,\eqref{eq:FDprop},
 the FD gauge is {\it not} a LC gauge. This is because the FD gauge vector\,\eqref{eq:FDprop}  does not transform as a four-vector under the Lorentz transformation, and hence the `dot product'
\begin{eqnarray}
n_{\rm FD}\cdot q=\left|q^0\right|+\left|\vect{q}\right|
\end{eqnarray}
is not Lorentz invariant. It is invariant under rotations while its magnitude transforms as $e^{\pm\eta}$ under boosts with rapidity $\pm\eta$ along the three-momentum direction~\cite{Chen:2022gxv}.

We further note that although the LC gauge expression of
the `polarization tensor', eq.\,(\ref{pol5}), has exactly the
same form with that of the FD gauge, except for its
particular orientation of the gauge vector $n^\mu$,
it is only in the FD gauge, the tensor can be expressed
in terms of the polarization vectors with definite
helicities, as shown explicitly in eqs.\,(\ref{eq:PFDmunu}) and\,(\ref{eq:PFDMN}).  
In general LC gauge, we can also express the polarization
tensor in terms of the three polarization vectors as
\begin{eqnarray}
 {P}_{\rm LC}{}^\mu_{\ \nu}
 &=&-g^{\mu}_{\ \nu}+\dfrac{n^\mu q_\nu+q^\mu n_\nu }{n\cdot q}
 \nonumber\\
 &=&
\sum_{\lambda=1,2}\epsilon^\mu(q,\lambda)\epsilon_\nu(q,\lambda)^\ast
\nonumber\\
&+&{\rm sgn}(q^2)\epsilon^\mu(q,3)\epsilon_\nu(q,3),
\end{eqnarray}
where the first two polarization vectors are orthogonal to both $q^\mu$ and
$n^\nu$,%
\begin{eqnarray}
\sum_{\lambda=1,2}\epsilon^\mu(q,\lambda)\epsilon_\nu(q,\lambda)^\ast
=
P_{\rm LC}{}^\mu_{\ \nu}-q^2\frac{n^\mu n_\nu}{(n\cdot q)^2},
\end{eqnarray}
 and the third one is
 \begin{eqnarray}
 \epsilon^\mu(q,3)=\frac{Q\,n^\mu}{n\cdot q}.
 \end{eqnarray}
However, none of the above three polarization vectors are helicity eigenstates.
It is only when the LC gauge vector $n^\mu$ is chosen along $n_{\rm FD}^\mu$, the three polarization vectors can be helicity eigenstates.

In the EW theory of the SM, there are triplets of
weak bosons, $W^1$, $W^2$ and $Z$, all of which have 
the same free Lagrangian as given by eq.\,(\ref{eq:L0}).
The mass term $m$ for $W^1$ and $W^2$ is the charged
weak boson mass, $m_W$, and the associated Goldstone
boson is $\pi^1$ and $\pi^2$, respectively, of the custodial $SU(2)$ triplets.
The charge eigenstates,
$W^\pm = (W^1 \mp i W^2)/\sqrt{2}$
and 
$\pi^\pm = (\pi^1 \mp i \pi^2)/\sqrt{2}$
are obtained from the pairs of real fields.
Consequently, the \nth{5} component of the charged weak
bosons are the Goldstone bosons of the same charge
in the LC (and in the FD) gauge.

\section{Sample results in QED and QCD}\label{sec:results}
As is clear from the explicit form of the general
LC gauge propagators of the SM gauge bosons,
the photon, the gluons, and the weak bosons
$W^\pm$ and $Z$, the only difference between
the LC gauge and the FD gauge is in the choice
of the LC gauge vector $n^\mu$.
The Feynman rules to obtain an arbitrary tree-level
scattering amplitudes are also identical among
all LC gauges, including the FD gauge.
This is in particular so, for the $ZZZZ$,
$WWZH$ and $WWAH$ verteices which are needed
to be introduced in ref.\!\cite{Chen:2022gxv}, in order to allow {\tt Madgraph} to generate necessary {\tt HELAS} subroutine calls and the corresponding Feynman diagrams.

Since all the {\tt HELAS} numerical codes have been
prepared for the FD gauge in refs.\!\cite{Hagiwara:2020tbx,Chen:2022gxv},
we introduce a switch in these codes\footnote{\url{https://madgraph.ipmu.jp/IPMU/Softwares/HELAS/index.html}} 
which allows us to fix $n^\mu$ globally,
independent of the gauge boson four momenta.
It is only by this minimal modification,
the {\tt HELAS} codes prepared for the FD gauge
can be used to generate tree-level helicity
amplitudes of an arbitrary SM processes.  

We note in passing that we choose not to introduce
the poarization vectors in the generic LC gauge,
because they are not helicity eigenstates,
as explained in the previous section.
When we need to calculate the scattering amplitudes
with external on-shell weak bosons, we use the
polarization vectors in the FD gauge, so that
we can obtain the helicty amplitudes even in
the generic LC gauge.  

In this section, we show sample results we obtain
by using our new programs for simple QED and QCD
processes, and also for an EW process in the next
section.
In all the samples, we compare results of a few
LC gauge vector directions against those of the
FD gauge and the Feynman gauge (for QED and QCD)
and against the Unitary gauge for the EW theory.

\subsection{QED}

\begin{figure}[h]
  \center
 \includegraphics[width=1\columnwidth]{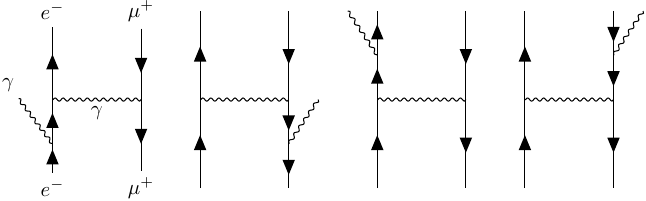}
\hspace*{0.5cm}(a)\hspace*{1.7cm}(b)\hspace*{1.9cm}(c)\hspace*{1.7cm}(d)
\caption{Feynman diagrams for $e^-\mu^+\to e^-\mu^+\gamma$ in QED\protect\footnotemark.}
\label{fig:diagram_em_ema}
\end{figure}

As the first simple example, we study the process
\begin{align}
  e^- + \mu^+ \to e^- + \mu^+ + \gamma
  \label{proc:em2emr}
\end{align}
in QED, whose Feynman diagrams are shown in Fig.\,\ref{fig:diagram_em_ema}.
It should be noted that each Feynman amplitudes of an arbitrary $2\to2$ processes in QED and QCD are gauge invariant among all covariant gauges and in LC gauges, because the photon and the gluon propagators are sandwiched by two conserved currents in both sides.
\footnote{Gauge dependence can appear in general axial gauge due to $n^\mu n^\nu$ term.}

\footnotetext{The Feynman diagrams in Figs.\ref{fig:diagram_em_ema},\ref{fig:diagram_gg_ggg}, and \ref{fig:diagram_aa_ww} are drawn with
{\tt TikZ}-{\tt FeynHand}\!\cite{Ellis:2016jkw,Dohse:2018vqo}.
}
%
\begin{figure*}[t]
\center
\includegraphics[width=1\textwidth]{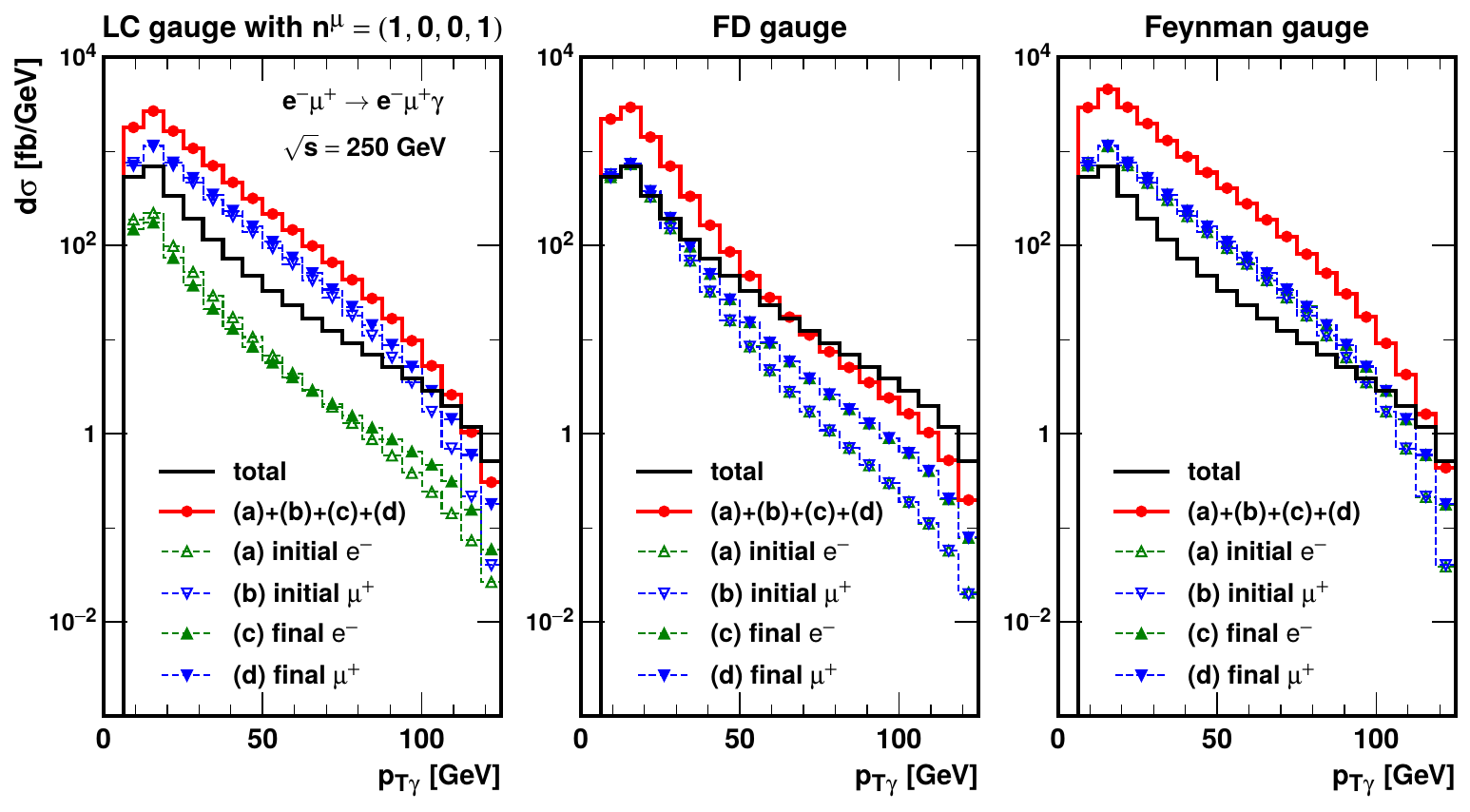}
\hspace*{1cm}(a)\hspace*{5.5cm}(b)\hspace*{5.5cm}(c)
\caption{$p_T$ distributions of the photon for $e^-\mu^+\to e^-\mu^+\gamma$
at $\sqrt{s}=250$~GeV in different gauges; 
(a) LC, (b) FD, and (c) Feynman gauges.}
\label{fig:pta}
\end{figure*}

\begin{figure*}[t]
\center
\includegraphics[width=1\textwidth]{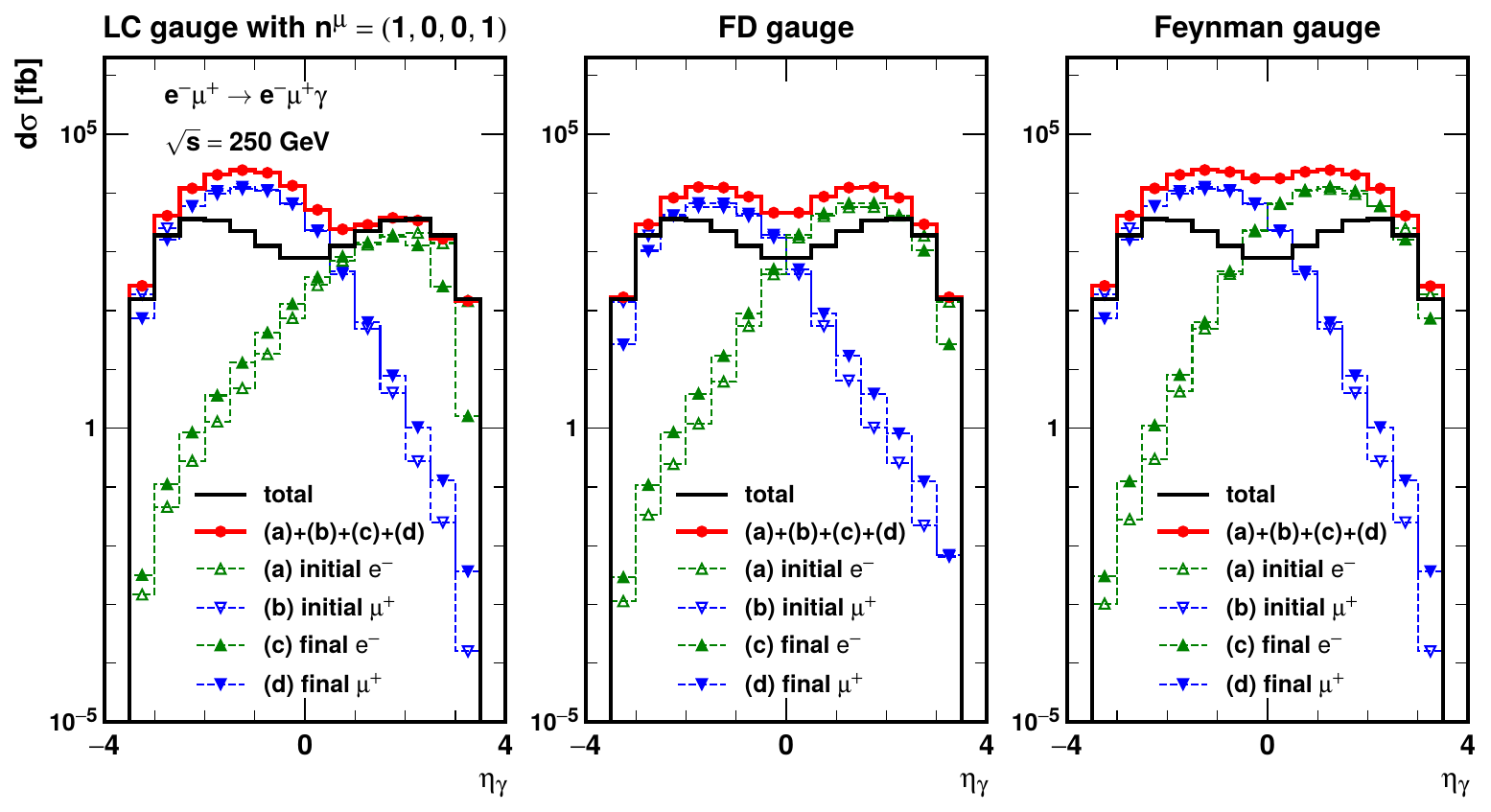}
\hspace*{1cm}(a)\hspace*{5.5cm}(b)\hspace*{5.5cm}(c)
\caption{Same as Fig.~\ref{fig:pta}, but for the photon rapidity $\eta_\gamma$ distributions.}
\label{fig:eta}
\end{figure*}

\begin{figure*}[t]
\center
\includegraphics[width=1\textwidth]{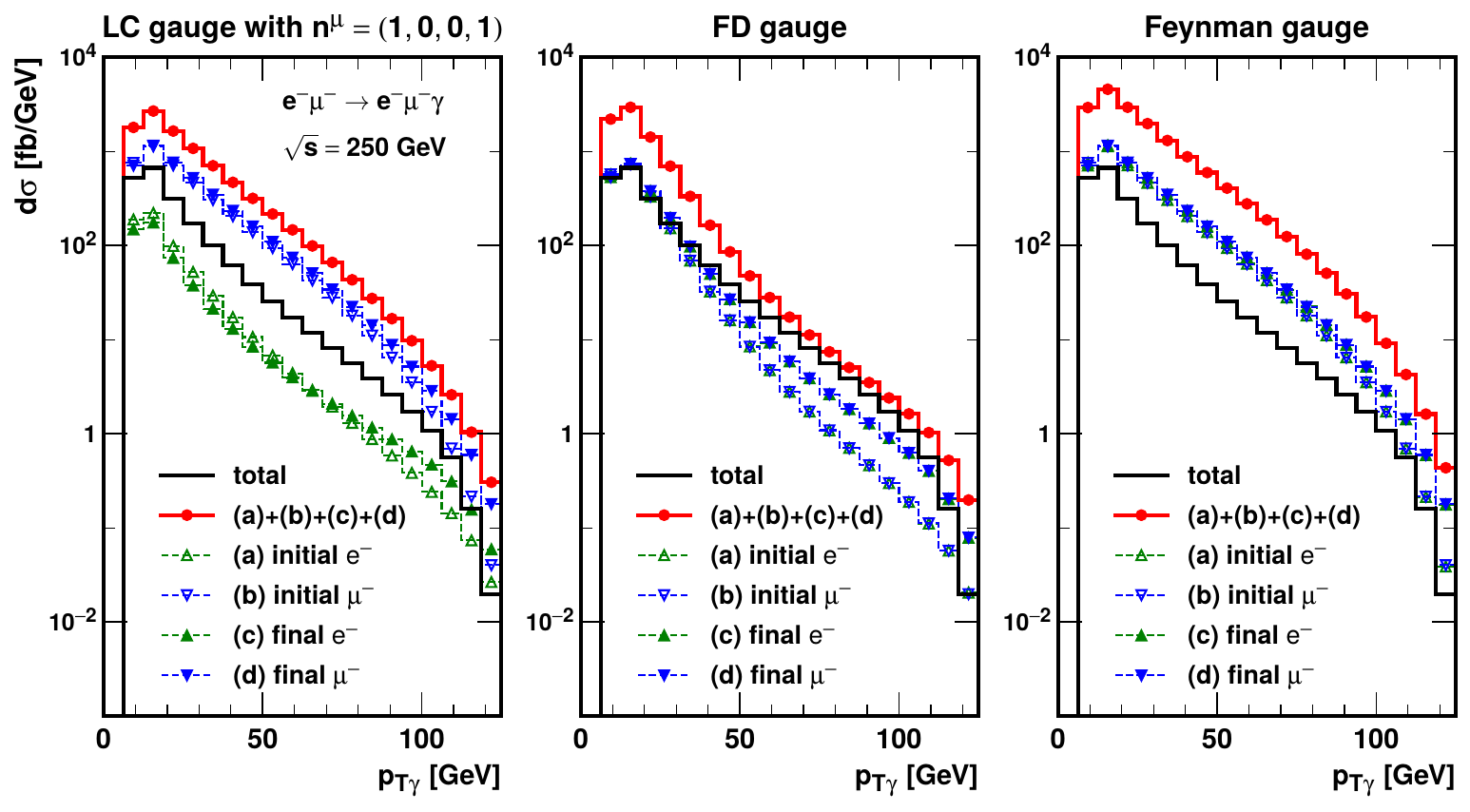}
\hspace*{1cm}(a)\hspace*{5.5cm}(b)\hspace*{5.5cm}(c)
\caption{
$p_T$ distributions of the photon for $e^-\mu^-\to e^-\mu^-\gamma$
at $\sqrt{s}=250$~GeV in different gauges; 
(a) LC, (b) FD, and (c) Feynman gauges.}
\label{fig:QEDpT250_emmm}
\end{figure*}
%

\begin{figure*}[t]
\center
\includegraphics[width=1\textwidth]{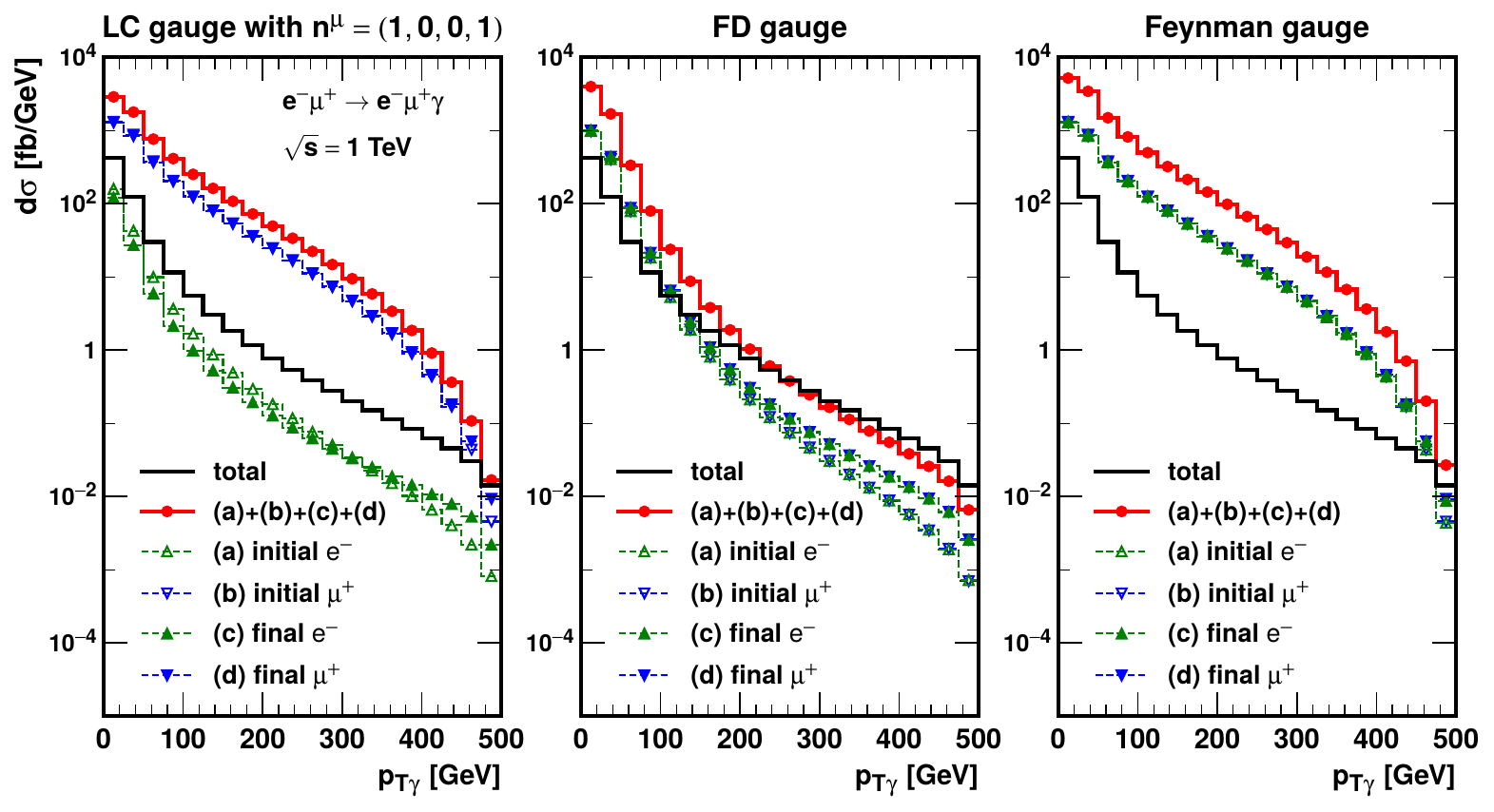}
\hspace*{1cm}(a)\hspace*{5.5cm}(b)\hspace*{5.5cm}(c)
\caption{
$p_T$ distributions of the photon for $e^-\mu^+\to e^-\mu^+\gamma$
at $\sqrt{s}=1$~TeV in different gauges; 
(a) LC, (b) FD, and (c) Feynman gauges.}
\label{fig:QEDpT1TeV_emmp}
\end{figure*}

We also note here that we show $e^- \mu^+$ collision
case rather than the $e^+ e^-$ or $\mu^+ \mu^-$
collisions simply to remove $\ell\bar{\ell}$ annihilation
contributions.
At high energies, at $\sqrt{s} = 250$~GeV and 1~TeV which
we study below, the annihilation processes give rise
to the so-called radiative return events, where
the final lepton pair comes dominantly from the
$Z$ boson decays, and the photon is emitted along
the beam momentum directions.
  Although our pure QED amplitudes do not have $Z$ boson contributions, the interference patterns reported in our study may be observable at future $e^+e^-$ or $\mu^+\mu^-$ colliders after imposing final state cut like $\left|m(\ell\bar{\ell})-m_Z\right|\gtrsim15~{\rm \Gamma}_Z$ to suppress $\ell\bar{\ell}$ annihilation contributions. 

We set the minimal kinematical cuts for the final-state particles
($i,j=e,\mu,\gamma$):
\begin{eqnarray}
  p_{T_i}&>&10~{\rm GeV},
  \label{subeq:pTcut}
  \n\\
  \Delta R_{ij}&=&\sqrt{(\eta_i-\eta_j)^2+(\phi_i-\phi_j)^2}>0.4.
    \label{eq:qedcuts}
  \end{eqnarray}

Note that the above final state cuts remove the region
of the phase space where the virtuality of the exchanged
photon is very small where the Feynman gauge (or in
any covariant gauge) amplitudes suffer from serious
numerical cancellation among interfering amplitudes\!\cite{Hagiwara:1990gk}.
In this paper, we study how the interference patterns
among individual Feynman diagram contributions differ
among a few selected LC gauges, Feynman gauge, in
contrast to the FD gauge.

In Fig.\,\ref{fig:pta}, we show the $p_T$ distributions
of the photon in the process eq.\,(\ref{proc:em2emr}) at $\sqrt{s}=250$~GeV. 
We compare three cases, (a) LC gauge with
\begin{eqnarray}
n^\mu = 
\begin{pmatrix}
1,& 0,& 0,&1
\end{pmatrix}, 
\label{eq:LCvector}
\end{eqnarray}
(b) FD, and (c) Feynman gauges.
Absolute value square of the four individual Feynman diagram
contribution is shown separately, and their sum is given by
the solid red histograms.
The physical distribution, obtained by summing all the
amplitudes before squaring, given by the thick black
histograms, are identical (gauge independent) as expected.

We observe from Fig.\,\ref{fig:pta}(c) that in the Feynman gauge,
the $p_T$ distribution from all the four diagrams are
larger than the physical distribution.
Contributions from all the four individual diagrams
have similar magnitudes at all $p_T$ range,
telling destructive interference among the
sub-amplitudes all the way up to $p_T\sim120$~GeV.  

In case of the FD gauge, in Fig.\,\ref{fig:pta}(b), we observe that the interference is 
destructive in the region
$p_T\lesssim60$ GeV, while constructive at
$p_T\gtrsim80$ GeV.
The destructive interference for low $p_T$ photon
is the property expected for soft-photon radiation, where the $\gamma$ emitted from the initial $e^-(\mu^+)$ and that from the final $e^-(\mu^+)$ cannot resolve initial and final leptons when the $\gamma$ $p_T$ is smaller than the final lepton $p_T$.
Indeed, at $p_T\lesssim30$~GeV, all four diagrams
give similar magnitude, and the sum of all four
diagrams give essentially the same magnitude
with individual contribution, as may be expected for
soft photons.

When comparing the above two cases, for the (b) FD gauge and the (c) Feynman gauge, against the Fig.\,\ref{fig:diagram_em_ema}(a) for the LC
gauge with the LC vector of eq.\,(\ref{eq:LCvector}),
we notice a striking asymmetry in the individual
Feynman diagram contribution between the
emissions from the $e^-$ leg (denoted by the
histogram with green triangles) and those 
from the $\mu^+$ leg (denoted by the histograms
with blue triangles).

Amplitudes with initial and final state emission
from the $e^-$ legs are suppressed, while those
from the $\mu^+$ legs are enhanced.
Because the sum of the two emission amplitudes,
(a) and (c) for the $e^-$ legs, and
(b) and (d) for the $\mu^+$ legs, are
separately gauge independent,
this means that the suppressed (a) and (c)
amplitudes interfere constructively,
and the enhanced (b) and (d) amplitudes
interfere destructively,
to obtain the same magnitudes.

The reason for the suppression of the amplitudes
(a) and (c) can be understood qualitatively as
follows.
The LC gauge vector\,(\ref{eq:LCvector}) is approximately the FD gauge vector for the $\gamma$ propagators in the diagrams Fig.\,\ref{fig:diagram_em_ema}(a) and (c), when the final $\mu^+$ has low $p_T$ satisfying the cut (\ref{subeq:pTcut}), so that the virtual photon
momentum is approximately along the initial $\mu^+$
momentum, which is beam energy times $(1,0,0,-1)$.
These amplitudes are then suppressed because the components which grow with the energy of the virtual gauge bosons are removed by the FD gauge prescription. 
On the contrary, in diagrams (b) and (d), whose magnitudes are large when the final $e^-$ has low $p_T$, the LC vector\,(\ref{eq:LCvector}) is opposite of the corresponding FD gauge vector, which should be $(1,0,0,-1)$ along the virtual $\gamma$ momentum direction. 
Rather than subtracting the terms which grow with the energy, these terms are essentially kept untouched to give contributions as big as those in the Feynman gauge, as shown in Fig.\,\ref{fig:pta}(c).

Although we do not show results in this report,
we examine another choices of the LC gauge vector,
  \begin{eqnarray}
n^\mu = 
\begin{pmatrix}
1, &0,& 0,& -1  
\end{pmatrix}.
\label{eq:LCv-1}
  \end{eqnarray}

As may be expected, the gauge vector\,\eqref{eq:LCv-1} reverses
the contributions of the photon emissions from
the $e^-$ legs, and those from the $\mu^+$ legs.
The magnitudes of the amplitudes (a) and (c) are now
enhanced, while those of the amplitudes (b) and (d)
are suppressed.

In Fig.\,\ref{fig:eta}, we show the rapidity ($\eta$) distribution
of the emitted photon in the process\,\eqref{proc:em2emr} at
$\sqrt{s} = 250$~GeV.
The same final state cuts\,\eqref{eq:qedcuts} are applied.
The FD gauge results in Fig.\,\ref{fig:eta}(b) show clearly that
the photons emitted from initial and final $e^-$ legs
tend to have positive rapidity, while those from
initial and final $\mu^+$ legs tend to have negative
rapidity, as naively expected from the parton shower
behavior.
They interfere destructively in the central region,
where soft (low $p_T$) photons from both initial
and final emissions can interfere.
Similar trends can be observed in the Feynman gauge
results shown in Fig.\,\ref{fig:eta}(c), but the degree of
destructive interference is of the order of $30\sim40$
rather than a factor of $5\sim7$ in case of the
FD gauge.

Finally, the LC gauge results for the gauge vector
of eq.\,\eqref{eq:LCvector} is shown in Fig.\,\ref{fig:eta}(a).
In this plot, we can clearly view the impact of
our choice of the gauge vector\,\eqref{eq:LCvector}.
Even though the amplitudes for the photon emitted
from the $e^-$ legs, Fig.\,\ref{fig:diagram_em_ema}(a) and (c), are
suppressed strongly, they still dominates at
very large positive $\eta$.
We can tell that the diagram Fig.\,\ref{fig:diagram_em_ema}(a) for the
photon emission from the initial $e^-$ dominates
at the highest rapidity region, $\eta\gtrsim 2.5$,
while the two diagrams (a) and (c) interfere
constructively in the medium positive rapidity
region, $1\lesssim\eta\lesssim 2.5$.
Destructive interference between the enhanced
amplitudes from the initial and final $\mu^+$
emissions, the Feynman diagrams of Figs.\,\ref{fig:diagram_em_ema}(b) and (d),
respectively, can be observed at all negative
rapidity.
The degree of cancellation is as big as in the
case of the Feynman gauge, shown in Fig.\,\ref{fig:eta}(c).

In Fig.\,\ref{fig:QEDpT250_emmm}, we show results for the process
\begin{eqnarray}
e^- + \mu^- \to e^- + \mu^- + \gamma
\label{proc:emmm}
\end{eqnarray}
at $\sqrt{s} = 250$~GeV.
Here, we want to compare Fig.\,\ref{fig:pta}(b) for the
$e^- \mu^+$ collisions and Fig.\,\ref{fig:QEDpT250_emmm}(b) for
the $e^- \mu^-$ collisions, both in the FD gauge.
Because individual Feynman diagram cannot tell
the charge of $\mu^-$ from that of $\mu^+$,
the absolute value square of each amplitude,
and hence their sum, are all the same between Fig.\,\ref{fig:pta}
and Fig.\,\ref{fig:QEDpT250_emmm} for all the gauges.
The physically observable distribution, given by
the black histogram, is common (gauge independent)
among the three gauges shown in Fig.\,\ref{fig:QEDpT250_emmm}, but it
is slightly different between Fig.\,\ref{fig:pta} for the
process\,\eqref{proc:em2emr} and Fig.\,\ref{fig:QEDpT250_emmm} for the process\,\eqref{proc:emmm}.

When we compare the two results in the FD gauge,
the physical distribution (in thick black histogram)
is bigger than the sum of the square of each diagram
(given by the thick red histogram) at large $p_T$,
$p_T \gtrsim70$~GeV, in Fig.\,\ref{fig:pta}(b) for the $e^- \mu^+$
collision process\,\eqref{proc:em2emr}.
On the other hand, in Fig.\,\ref{fig:QEDpT250_emmm}(b) for the $e^- \mu^-$
collision process\,\eqref{proc:emmm}, the physical cross section
is never greater than the sum of the square of
individual Feynman amplitude at all $p_T$.
Therefore, at large $p_T$, $p_T \gtrsim70$~GeV we observe constructive interference among
the Feynman amplitudes for the $e^- \mu^+$ collision
process\,\eqref{proc:em2emr}, whereas destructive interference in
case of the $e^- \mu^-$ collision.

This observation is in agreement with what we can
expect from the null-radiation zone theorem\!\cite{Brown:1982xx},
which tells that the photon emission is absent
when the following identity holds
  \begin{eqnarray}
\sum_{\rm All\,charged\,particles} Q_j/(p_j \cdot k) = 0 ,
\end{eqnarray}
where $Q_j$ and $p_j^\mu$ are the charge and four momentum
of $j^{\rm th}$ charged particle, and $k^\mu$ is the photon
four momentum.
In ref.\!\cite{Hagiwara:2012xj} the above identity has been applied to 
the scattering process $ab \to cd\gamma$ and the
following conditions are obtained:
\begin{subequations}
\begin{align}
&Q_a = Q_b = Q_c = Q_d, \\
&\eta(\gamma) = \eta(a + b) = 0 , \\
&p_T(c) = p_T(d),      
\end{align}
\end{subequations}
in the $a+b$ collision c.m. frame.
  Here, $\eta(\gamma)=\eta_\gamma$ and $\eta(a+b)$ are the rapidity of the photon and the colliding $a+b$ system.
The theorem tells that in the c.m. frame of the $e^- \mu^-$ collision
process\,\eqref{proc:emmm}, there should be a perfect destructive
interference among the amplitudes when the final $e^-$
and $\mu^-$ have the same $p_T$ and the rapidity of
the photon is zero.
We can observe that these conditions are relatively easy to be satisfied when the photon $p_T$ is large, because large $p_T$ photons tend to have small rapidity, and the recoil particles $e^-$ and $\mu^-$ tend to have $p_T$ of the same order of magnitude.

The destructive interference associated with the
presence of the null radiation zone at high photon
$p_T$ for the $e^- \mu^-$ collision process\,\eqref{proc:emmm}
implies that strong constructive interference
should be expected for the $e^- \mu^+$ collision
process\,\eqref{proc:em2emr} at the same kinematical configuration\!\cite{Hagiwara:2012xj,Hagiwara:2012we}.
When individual Feynman amplitudes are expressed
in the FD gauge, this physical picture can be
manifestly observed, as shown by the high $p_T$
behavior shown in Figs.\,\ref{fig:pta}(b) and \,\ref{fig:QEDpT250_emmm}(b).
Since the null radiation theorem accounts only for
classical charged currents, both the destructive
and constructive interference patterns are independent
of the lepton helicities.

In contrast, individual Feynman amplitudes are not enlightening in studying such interference patterns, as can be observed by comparing Figs.\,\ref{fig:pta}(c) and\,\ref{fig:QEDpT250_emmm}(c) for the Feynman gauge, and by comparing Fig.\,\ref{fig:pta}(a) and\,\ref{fig:QEDpT250_emmm}(a) for the LC gauge.

Although we study only the interference patterns
for soft-photon emissions and those associated
with the null radiation theorem in this report,
the FD gauge amplitudes may tell us more about
physics of interference among Feynman amplitudes.

Before closing the subsection on the QED amplitudes, let us show in Fig.\,\ref{fig:QEDpT1TeV_emmp} the photon $p_T$ distribution of the process\,\eqref{proc:em2emr} $e^-\mu^+\to e^-\mu^+\gamma$ at $\sqrt{s}=1$ TeV. As in Fig.\,\ref{fig:pta}, the three figures are for (a) LC gauge with $n^\mu=(1,0,0,1)$, (b) FD gauge, and (c) Feynman gauge. 

The destructive interference at low $p_T$, as expected for soft photon, and the constructive interference at high $p_T$, consistent with prediction of the classical radiation zero theorem~\cite{Brown:1982xx}, are clearly seen in the FD gauge, in Fig.\,\ref{fig:QEDpT1TeV_emmp}(b). 
Although both LC gauge and Feynman gauge results don't give us such insights, we can tell by comparing Fig.\,\ref{fig:QEDpT1TeV_emmp}(a) and (c) that the magnitude of physical enhancement of the amplitudes which are not suppressed by the LC gauge vector of $n^\mu=(1,0,0,1)$, Figs.\,\ref{fig:diagram_em_ema}(b) and (d), are as large as those of the Feynman gauge. The difference is that in case of the Feynman gauge, Fig.\,\ref{fig:QEDpT1TeV_emmp}(c), all four amplitude squares are large, and hence their sum given by the red histogram is about twice as large as the corresponding sum in Fig.\,\ref{fig:QEDpT1TeV_emmp}(a) for the LC gauge.
%

\subsection{QCD}
\begin{figure}[h]
  \center
\includegraphics[width=0.9\columnwidth]{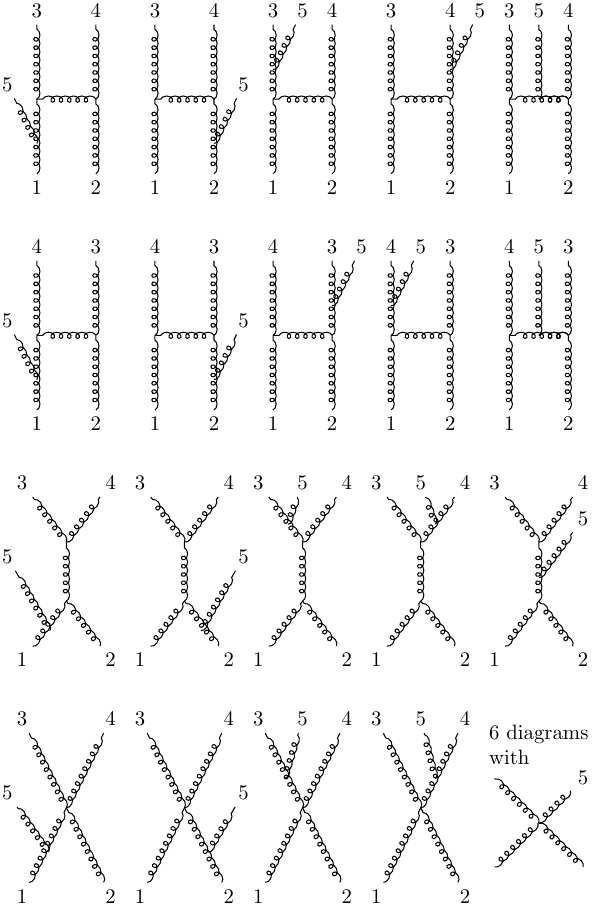}
\hspace*{0.3cm}(a)\hspace*{1.2cm}(b)\hspace*{1.2cm}(c)\hspace*{1.2cm}(d)\hspace*{1.2cm}(e)
\caption{Feynman diagrams for $gg\to ggg$.}
\label{fig:diagram_gg_ggg}
\end{figure}

In this subsection, we study LC gauge amplitudes in QCD $2\to3$ processes, and compare with the FD gauge and the Feynman gauge predictions\!\cite{Hagiwara:2020tbx}. For the quark-quark scattering process $ud\to udg$  \!\cite{Hagiwara:2020tbx}, we find that it gives more or less similar results to the QED example of $e^-\mu^\pm\to e^-\mu^\pm\gamma$ shown in the previous subsection. Therefore, we only present our studies on the process 
\begin{align}
  g(p_1) + g(p_2) \to g(p_3) + g(p_4) + g(p_5). 
  \label{proc:gg2ggg}
\end{align}
The Feynman diagrams are depicted in Fig.~\ref{fig:diagram_gg_ggg},
where we omit six diagrams with the four-point vertex
in the column (e).
The diagrams are grouped into 5 types along columns. 
In the first column, type (a), the gluon (5) is emitted from $g(1)$.
Likewise, $g(5)$ is emitted from $g(2)$ in the second column (b), a virtual gluon splits into $g(3)$ and $g(5)$ in (c), or into $g(4)$ and $g(5)$ in (d).
In the diagrams of type (e), $g(5)$ is emitted from an exchanged virtual gluon or from the four-point vertex.
%

 %

\begin{figure*}[h]
  \center
\includegraphics[width=1\textwidth]{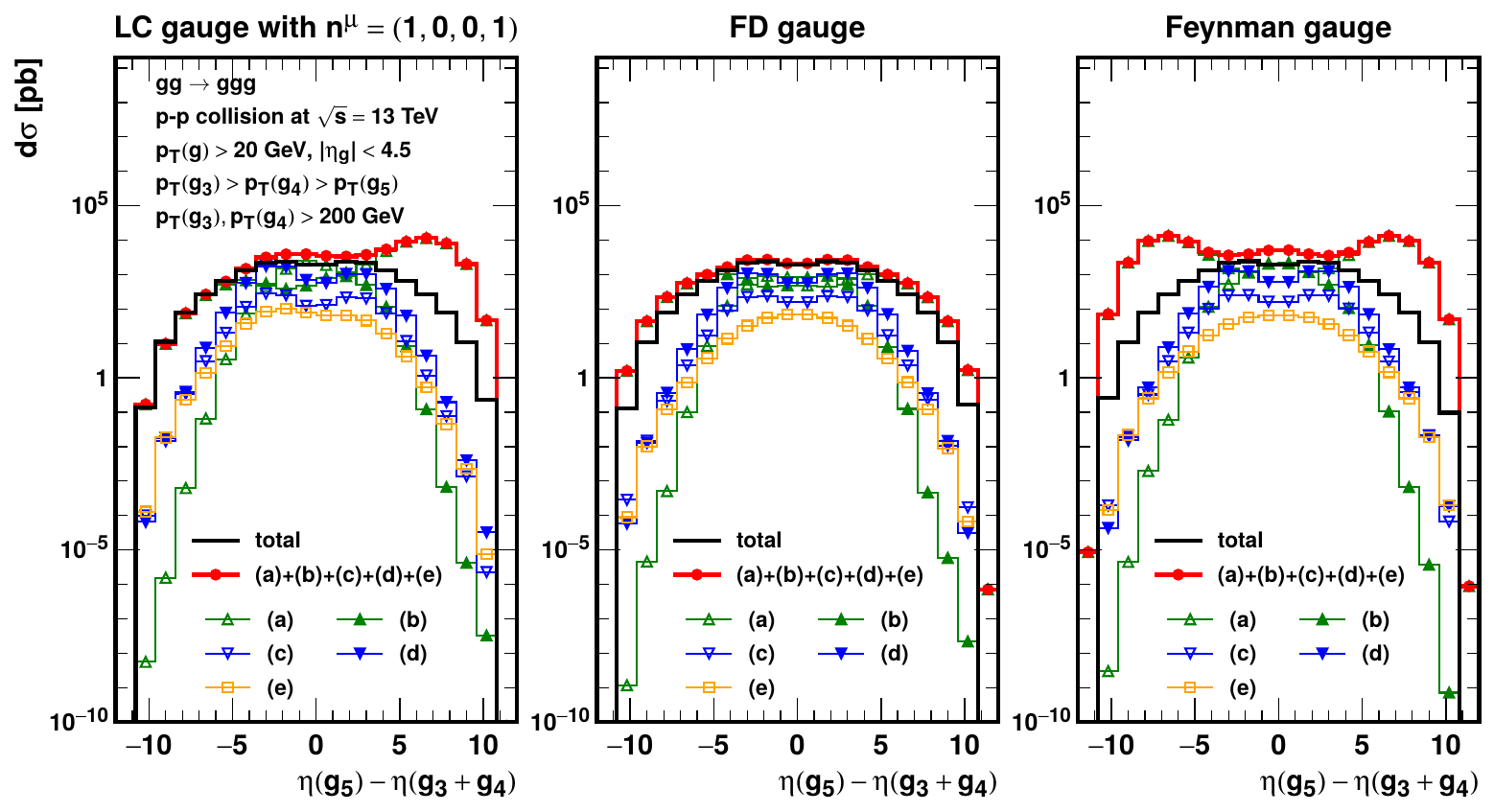}
\hspace*{1cm}(a)\hspace*{5.5cm}(b)\hspace*{5.5cm}(c)
\caption{ 
Distributions of the rapidity difference,
$\eta_5-\eta_{34}$, where $\eta_{34}$ is the rapidity of the hard di-jet system of two gluons, $g(p_3)$ and $g(p_4)$, in $gg\to ggg$ subprocess contribution to $pp$ collisions at $\sqrt{s}=13$ TeV.
 They are calculated in (a) LC gauge with $n^\mu=(1,0,0,1)$,
  (b) FD gauge, and (c) Feynman gauge.   
  The black solid histogram gives the cross section
  predicted by QCD, and hence agrees among the three
  plots, while the red solid histogram shows the sum
  of the absolute value square of each Feynman amplitude.
  Sub-contributions from the 5 groups of Feynman
  diagrams in Fig.\,\ref{fig:diagram_gg_ggg} are also shown.
}
\label{fig:eta534}
\end{figure*}
 %
 
\begin{figure*}[h]
\center
\includegraphics[width=1\textwidth]{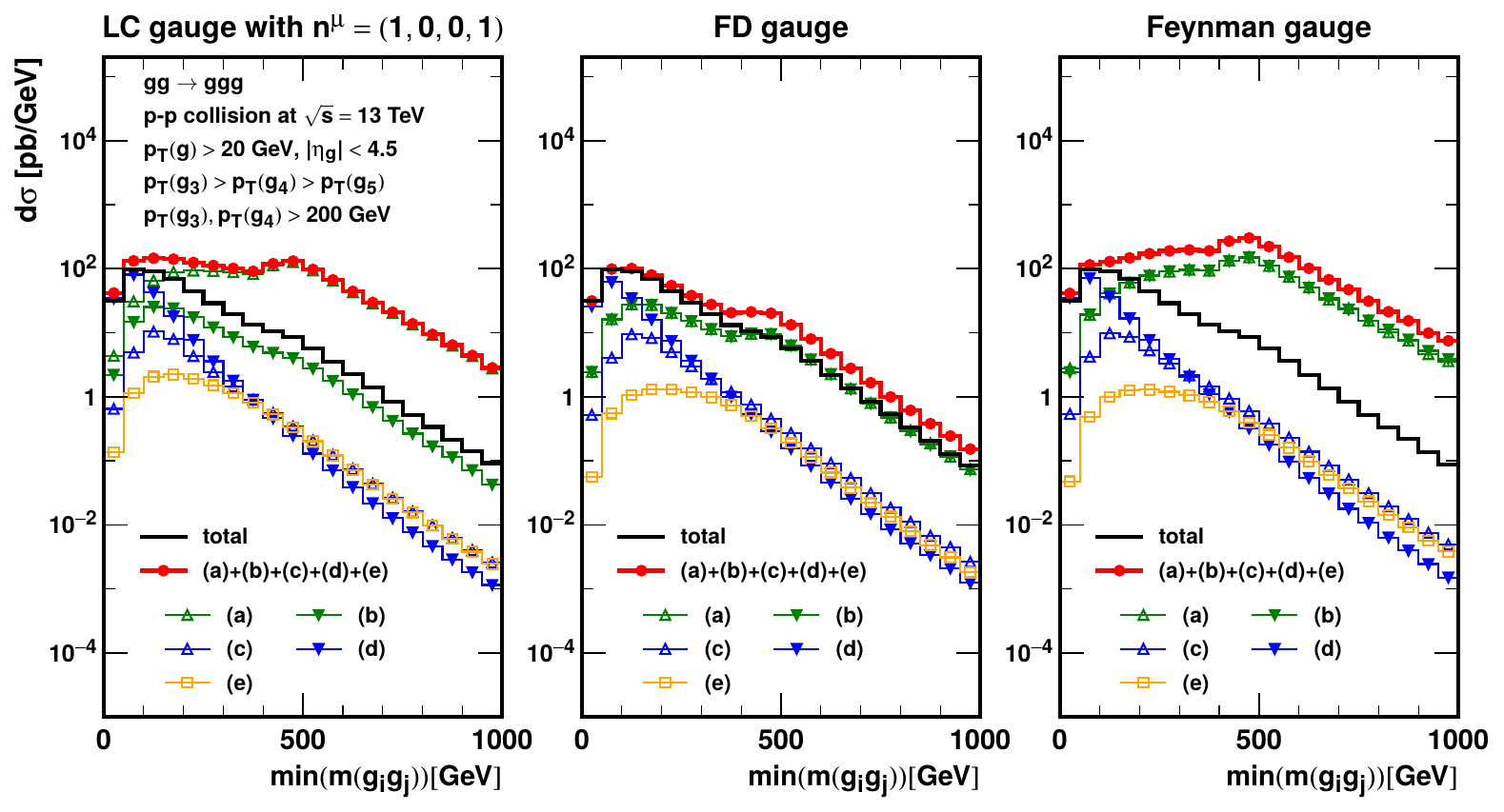}
\hspace*{1cm}(a)\hspace*{5.5cm}(b)\hspace*{5.5cm}(c)
\caption{
The same as Fig.\,\ref{fig:eta534}, but for the distribution of ${\rm min}(m_{ij})$, where $m_{ij}$ are the invariant mass of the final state di-gluons among $g(p_3)$, $g(p_4)$ and $g(p_5)$.
 }
\label{fig:gg_ggg}
\end{figure*}

Event samples of $gg\to ggg$ are generated for $pp$ collisions
at $\sqrt{s}=13$~TeV, by using the gluon PDF of ref.~\cite{Pumplin:2002vw}.
We impose the following final state cuts:
\begin{subequations}
\begin{align}
 &p_{Ti}>20~{\rm GeV},
 \label{eq:pTcut}
 \\ 
   &|\eta_i| < 4.5,
      \label{eq:etacut}
   \\
 &\Delta R_{ij}=\sqrt{(\eta_{i}-\eta_{j})^{2} + (\phi_{i} - \phi_{j})^{2}}>0.4,
 \label{eq:delrcut}
\end{align}
\label{eq:cuts}
\end{subequations}
for $i,j=3,4,5$ for the final state gluons in eq.\,\eqref{proc:gg2ggg}.  
We set the scale of the colliding gluon PDF at $Q=200$ GeV, and fix the QCD coupling at $\alpha_S=\alpha_S(200~{\rm GeV})=0.106$.

The initial gluon $g(p_1)$ has momentum along the positive $z$-axis, while $g(p_2)$ momentum is along the negative $z$-axis. The three final state gluons are identified by $p_T$ ordering,
\begin{eqnarray}
p_{T3}>p_{T4}>p_{T5},
\label{eq:ptorder}        
\end{eqnarray}
and we introduce a hard scale of two high $p_T$ jet production by requiring
\begin{eqnarray}
p_{T3},p_{T4}>200~{\rm GeV}. 
\label{eq:200GeVcut}
\end{eqnarray}
With the above setups, we can study the soft gluon radiation patterns by the $p_5$ dependences of the lowest $p_T$ jet. 

Thanks to the selection cut of\,\eqref{eq:200GeVcut}, we can interpret the 5 groups of the Feynman diagrams in Fig.\,\ref{fig:diagram_gg_ggg} representing the 5 types of parton shower histories: 
the 4 diagrams in the column (a) contribute to initial state
radiation from $g(1)$ before the hard $2\to2$ scattering process,
the 4 diagrams in the column (b) contribute to initial state
radiation from $g(2)$,
while those in the column (c) contribute to the final state
radiation from the hard jet $g(3)$,
those in the column (d) contribute to the final state radiation
from the hard jet $g(4)$,
whereas the 9 diagrams in the column (e) contribute to
soft gluon $g(5)$ emission from the hard $2\to2$ scattering process as a whole.

In Fig.\,\ref{fig:eta534}, we show the boost invariant distribution of the rapidity
difference
\begin{eqnarray}
\eta_5 -\eta_{34},   
\end{eqnarray}
where $\eta_{34}$ denotes the rapidity of the
hard scattering object, the system of $g(3)$ and $g(4)$,
whose four momentum is $p_3+p_4$.
We show the results obtained in three different gauges, in (a) LC gauge with $n^\mu=(1,0,0,1)$, (b) FD gauge , and (c) Feynman gauge. 

There are 25 Feynman diagrams for the process $gg\to ggg$ as
depicted in Fig.\,\ref{fig:diagram_gg_ggg}.
The physical cross section depends on the squared of the
sum of all the 25 Feynman amplitudes,
\begin{eqnarray}
\left| \sum_{k=1}^{25} {\cal M}_k \right|^2 ,              
\end{eqnarray}
which are given by solid black histograms.
As expected, they agree exactly among the three gauges.
What we show by solid red histograms are the contribution of
\begin{eqnarray}
\sum_{k=1}^{25} \left|{\cal M}_k\right|^2,     
\end{eqnarray}
that is, the sum of the absolute value squares of all 25 individual Feynman 
amplitudes.
Where the red histogram is significantly above the
black histogram, we can tell that subtle cancellation
among Feynman amplitudes takes place.
Furthermore, contribution to the red histograms are
divided into 5 groups, denoted by (a) to (e), according
to those in the 5 columns, respectively, in Fig.\,\ref{fig:diagram_gg_ggg}. 

As expected,
we find that there is no subtle gauge theory cancellation
in the FD gauge, as shown in Fig.\,\ref{fig:eta534}(b), where the red-solid histogram 
is almost degenerate with the observable black-solid
histogram in most region of the rapidity difference, $|\eta_5-\eta_{34}|\lesssim8$.
The large positive region of the rapidity difference is
dominated by the initial state radiation from $g(1)$, the diagram group (a),
whereas the large negative region is dominated
by the initial state radiation from $g(2)$, the diagram group (b).
The central region is populated by
gluon emission from the final state hard gluons,
$g(3)$ or $g(4)$.
Soft gluon emission from the hard process,
depicted by orange histogram, populates in
the central region but is always subdominant.

In comparison, in both the LC gauge and in the Feynman gauge,
the red solid histogram is significantly above the
black solid histogram in a broad region of
the rapidity difference.
The contributions from initial state radiations,
the Feynman diagram groups (a) and (b) in Fig.\,\ref{fig:diagram_gg_ggg}
dominate the Feynman gauge distribution in Fig.\,\ref{fig:eta534}(c),
while in Fig.\,\ref{fig:eta534}(a) for LC gauge with $n^\mu=(1,0,0,1)$, only the initial state radiation from
$g(1)$ gets large.

Cross examination of the three plots in Fig.\,\ref{fig:eta534} reveals the followings. 
The degree of cancellation, or the ratio of the red and black histograms grow sharply at $\eta_5-\eta_{34}\gtrsim4$ both in the LC gauge, Fig.\,\ref{fig:eta534}(a), and in the Feynman gauge, Fig.\,\ref{fig:eta534}(c).
Although the red histogram is symmetric about $\eta_5-\eta_{34}=0$ in both FD and Feynman gauges, we observe no significant cancellation in the region of  $\eta_5-\eta_{34}<-4$ in case of the LC gauge with $n^\mu=(1,0,0,1)$, where the red and black histograms are almost degenerate. Looking back to the FD gauge results shown in Fig.\,\ref{fig:eta534}(b), we notice that there is a signal of non-negligible destructive interference at very large rapidity difference, $|\eta_5-\eta_{34}|\gtrsim8$.
If the FD gauge amplitudes are physical, the destructive interference observed in Fig.\,\ref{fig:eta534}(b) at large $|\eta_5-\eta_{34}|$ should have physical interpretation, whereas the absence of cancellation at $\eta_5-\eta_{34}<-8$ in Fig.\,\ref{fig:eta534}(a) should be a gauge artefact, or a specifically arranged benefit\!\cite{Nagy:2007ty,Nagy:2014mqa}, of the particular LC gauge vector.
We will come back to this problem below, after studying the di-jet invariant mass distribution.

In Fig.\,\ref{fig:gg_ggg}, we show the distribution of the
minimum of the three di-jet masses in the final state,
\begin{eqnarray}
{\rm min}(m_{ij})={\rm min}\{ m_{34}, m_{35}, m_{45} \},            
\end{eqnarray}
where $m_{ij}$ denotes the invariant mass of the final $g(p_i)$ and
$g(p_j)$ system.
The three gauge cases are shown again for
the LC gauge with $n^\mu=(1,0,0,1)$ in Fig.\,\ref{fig:gg_ggg}(a),
the FD gauge in Fig.\,\ref{fig:gg_ggg}(b),
and the Feynman gauge in Fig.\,\ref{fig:gg_ggg}(c).

As expected, the red histogram is almost degenerate
with the black histogram in Fig.\,\ref{fig:gg_ggg}(b) for the FD gauge, up to ${\rm min}(m_{ij})\simeq350$ GeV,
which confirms that there is no significant
cancellation in the FD gauge at small ${\rm min}(m_{ij})$.
At smallest mass region, the contribution of the diagrams
in (d), final state emission from $g(4)$ dominates,
the contributions from the initial state radiations, the diagrams (a) and (b) are sub-dominant, followed by the final state emission from $g(3)$.
Soft radiation contribution from the group (e)
is small at all mass range.

In contrast, the red histogram
increases with the di-jet mass up to $\sim500$ GeV in case of the Feynman gauge in Fig.\,\ref{fig:gg_ggg}(c), whereas in case of the LC gauge with $n^\mu=(1,0,0,1)$ in Fig.\,\ref{fig:gg_ggg}(a), the red histogram stays about constant up to $\sim500$ GeV. 
In case of the Feynman gauge, the large
unphysical contribution is dominated by the two
initial state radiation contributions, the diagram
groups (a) from $g(1)$ and those (b) from $g(2)$.
Their contributions, depicted by green triangles
and by green up-side down triangles, respectively,
are identical.
On the other hand, in the LC gauge with $n^\mu=(1,0,0,1)$,
the contribution from the initial radiation from
$g(2)$, depicted by green up-side-down triangles,
is suppressed.
The huge unphysical contribution at large mass
region is hence dominated by initial state radiation
from $g(1)$, denoted by green triangles.
As in our QED studies, this shows that the virtual
gluon propagator in the initial state radiation
diagrams in the column (b) in Fig.\,\ref{fig:diagram_gg_ggg}, is suppressed
for $n^\mu=(1,0,0,1)$, because it is collinear to
the FD gauge vector for the same propagator when
emitted gluon $g(5)$ is soft.
The role of the initial state radiation from
$g(1)$ and $g(2)$ is interchanged if we choose
the LC gauge vector as $n^\mu=(1,0,0,-1)$.

In all gauges, the smallest mass region is dominated
by the final state emission amplitudes, those of
the group (d), as expected.
The unphysical gauge cancellation becomes significant
above the invariant mass around 200~GeV both for
the (a) LC gauge, and for the (c) Feynman gauge.  
%

\begin{figure*}[h]
  \center
\includegraphics[width=1\textwidth]{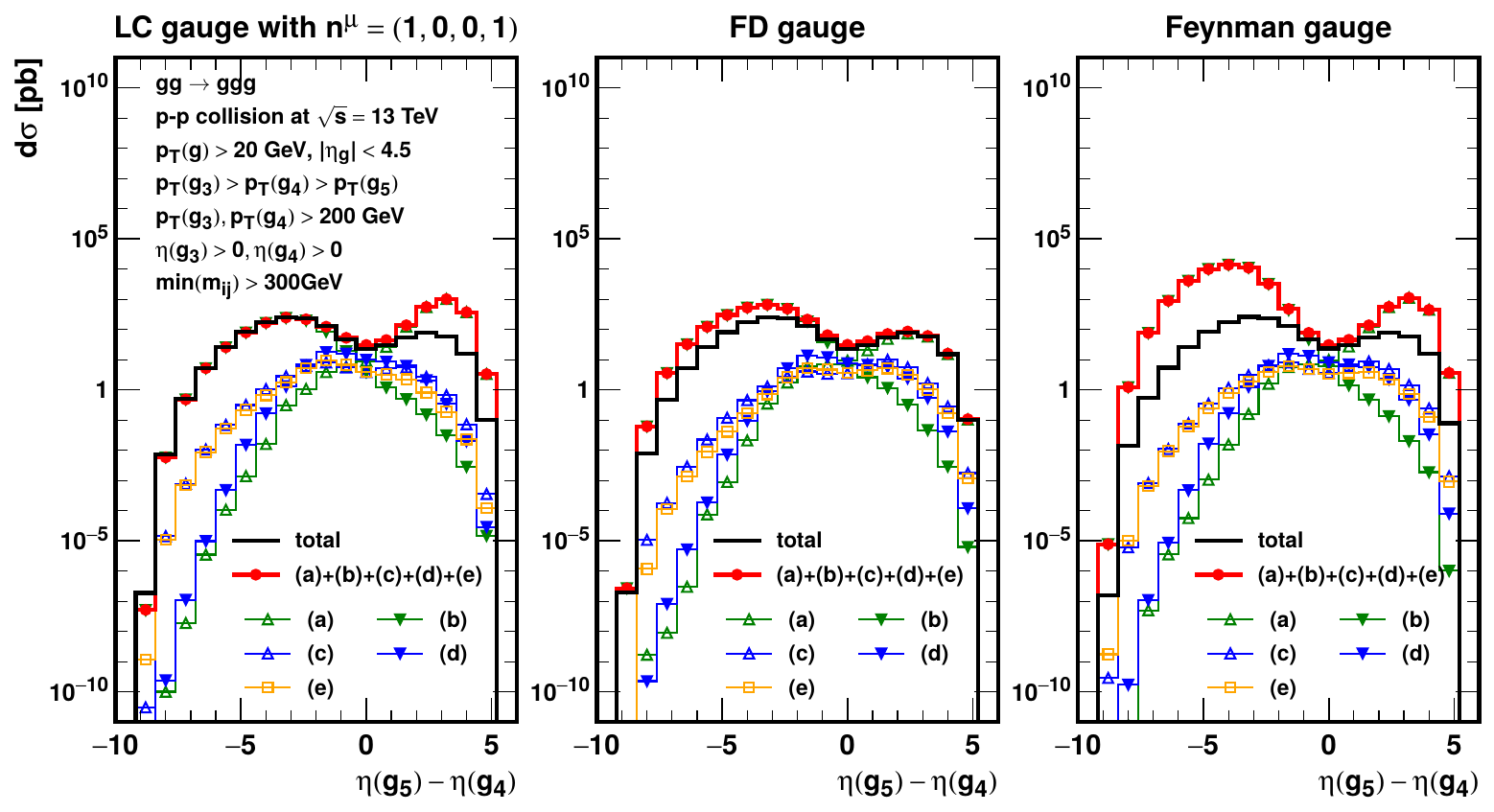}
\hspace*{1cm}(a)\hspace*{5.5cm}(b)\hspace*{5.5cm}(c)
\caption{The same as Fig.\,\ref{fig:eta534} but for 
distributions of the rapidity difference $\eta_5-\eta_4$, in $gg\to ggg$ subprocess contribution to $pp$ collisions at $\sqrt{s}$=13 TeV when ${\rm min}(m_{ij})>300$ GeV and $\eta_3,\eta_4>0$.
}
\label{fig:eta54}
\end{figure*}

 Let us now examine the distinct behavior of the red histogram around the medium di-jet mass region in Fig.\,\ref{fig:gg_ggg}(b) for the FD gauge. The red histogram stays almost constant for the three bins, in the range  
 \begin{eqnarray}
 350~{\rm GeV}\lesssim~{\rm min}(m_{ij})\lesssim500~{\rm GeV}
 \label{eq:min_mij_range350500}
 \end{eqnarray}
 In addition, significant destructive interference is observed in the region ${\rm min}(m_{ij})\gtrsim400$ GeV. 
 If the FD gauge amplitudes are all physical, we will be able to identify the cause of the peculiar behavior of the red histogram. We can also find the origin of the destructive interference which makes the black histogram smooth. 

 We first note that the kink of the red histogram at ${\rm min}(m_{ij})\simeq400$ GeV is observed in all gauges, which is dominated by the diagrams with initial state radiations, (a) and/or (b). 
 Since non-analytic behavior of the cross sections can only appear through our kinematical cuts, as depicted by eqs.\,\eqref{eq:cuts} and \eqref{eq:200GeVcut}, we examine the 3-jet configuration which is constrained by the cuts. 
 We find that the region around
 \begin{eqnarray}
 {\rm min}(m_{ij})\simeq350~{\rm GeV}
 \end{eqnarray}
 is dominated by the events where 
 \begin{eqnarray}
 \eta_3\simeq\eta_4\simeq\pm4.5.
 \end{eqnarray}
 That is, both hard jets with $p_T>$ 200GeV, eq.\,\eqref{eq:200GeVcut}, are at the boundary of our rapidity cuts, eq.\,\eqref{eq:etacut}. 
 The soft gluon $g(5)$ satisfying $p_T>20$ GeV, eq.\,\eqref{eq:pTcut}, can combine with either $g(3)$ or $g(4)$ to make ${\rm min}(m_{ij})$, while satisfying the jet separation cut of\,\eqref{eq:delrcut}. 
 For instance, in the positive rapidity region, the kinematical configuration of 
 \begin{subequations}
 \begin{align}
 p_3^\mu&=220~{\rm GeV}(\cosh4.5,~1,~0,~\sinh4.5),
 \\
  p_4^\mu&=200~{\rm GeV}(\cosh4.5,~-1,~0,~\sinh4.5),
 \\
   p_5^\mu&=20~{\rm GeV}(\cosh4.1,~-1,~0,~\sinh4.1),
 \end{align}
 \end{subequations}
 satisfies the $p_T$ balance and all the selection cuts of eqs.\,\eqref{eq:cuts}-\eqref{eq:200GeVcut}. We find
 \begin{eqnarray}
 {\rm min}(m_{ij})=m_{45}\simeq333~{\rm GeV}.
 \end{eqnarray}
 When $g(5)$ has azimuthal angles and rapidities satisfying the separation conditions of eq.\,\eqref{eq:delrcut}, the minimum di-jet mass stays in the region of $350~{\rm GeV}\lesssim{\rm min}(m_{ij})\lesssim500$ GeV where we observe a plateau in the red histogram of Fig.\,\ref{fig:gg_ggg}(b). 
 Most importantly, the rapidity of the two hard jets, $g(3)$ and $g(4)$, stay around their edge values of $\pm4.5$ allowed by our event selection cut \,\eqref{eq:etacut}.
 
 We now understand the cause of the destructive interference as observed in the region of 
 \eqref{eq:min_mij_range350500} as
 in Fig.\,\ref{fig:gg_ggg}(b).
  The soft gluon $g(5)$ cannot have larger rapidity than the hard jets, $g(3)$ and $g(4)$, because their rapidities are at the maximum value accepted by our selection cuts. Almost all $g(5)$ rapidity should hence satisfy 
 \begin{eqnarray}
 \eta_5<\eta_3,\eta_4,
 \end{eqnarray}
 in the positive rapidity region, and hence the soft gluon is emitted with larger polar angle than the hard jets. 
 The destructive interference we observe in Fig.\,\ref{fig:gg_ggg}(b) should hence be due to the same physics mechanism that leads to the angular ordering\cite{Dokshitzer:1982fh,Bassetto:1982ma, Ellis:1986bv} of QCD jet evolutions.
 
 As a confirmation of our observation, we show in Fig.\,\ref{fig:eta54} the distribution of the rapidity difference
 \begin{eqnarray}
 \eta_5-\eta_4
 \end{eqnarray}
  for those events which satisfy 
  \begin{eqnarray}
  {\rm min}(m_{ij})>300~{\rm GeV}
  \label{eq:min_mij_300}
  \end{eqnarray}
  and for the hemisphere of $\eta_3,\eta_4>0$.
  The destructive interference in the region of $\eta_5-\eta_4<0$ is clearly seen by comparing the red and black histograms in Fig.\,\ref{fig:eta54}(b) for the FD gauge. 
  We observe no significant interference in the region of $\eta_5-\eta_4>0$, where the soft gluon $g(5)$ has smaller polar angle than the hard jet. 
 
 We note that this cancellation of large angle soft gluon
emission is what we observe in the inclusive rapidity
difference distribution in Fig.\,\ref{fig:eta534}(b) in the region of large $\left|\eta_5-\eta_{34}\right|$.
The destructive interference effect is enhanced in
Fig.\,\ref{fig:eta54}(b) because we select those events with
min$( m_{ij} ) > 300$~GeV in eq.\,\eqref{eq:min_mij_300}, where the effect is
found to be large in the di-jet mass distribution
of Fig.\,\ref{fig:gg_ggg}(b).

In contrast to the FD gauge case we study above,
the Feynman gauge results in Fig.\,\ref{fig:eta54}(c) show that
significant destructive interference is found both
in the positive and negative rapidity ordering,
whereas the LC gauge results for
$n^\mu=(1,0,0,1)$ in Fig.\,\ref{fig:eta54}(a) show strong desctructive interference
at $\eta_5>\eta_4$ but no significant interference effects
at $\eta_5<\eta_4$.
Although detailed study of the LC gauge vector dependence of the gluon emission patterns in QCD amplitudes are beyond the scope of this report, we note here that the vector $n^\mu=(1,0,0,1)$ is along the opposite momentum direction of the initial gluon $g(2)$, which is regarded as a parent of the gluons emitted in the negative rapidity region.
 Likewise, if we choose the LC vector as  $n^\mu=(1,0,0,-1)$, the subtle gauge theory cancellation observed
in the region of large positive rapidity
difference in Figs.\,\ref{fig:eta534}(a) and \,\ref{fig:eta54}(a) disappears,
and re-appears in the negative rapidity
difference regions.
Therefore, if we choose $n^\mu=(1,0,0,1)$ for
those events with $\eta_5-\eta_{34}>0$,
while $n^\mu=(1,0,0,-1)$ for
events with $\eta_5-\eta_{34}<0$,
we can obtain the exact QCD matrix elements
with almost no cancellation for the
$gg \to ggg$ subprocess. 


\section{Studies in an EW process}\label{sec:EW}

\begin{figure}[t]
  \center
 \includegraphics[width=1\columnwidth]{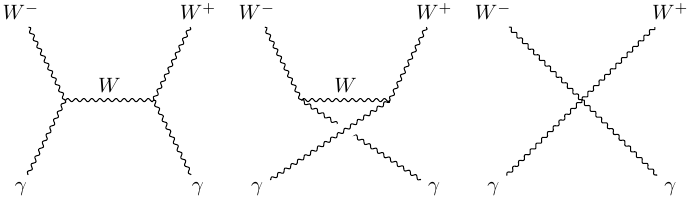}
(a)\hspace*{2.65cm}(b)\hspace*{2.65cm}(c)
\caption{Feynman diagrams for $\gamma\gamma\to W^-W^+$.}
\label{fig:diagram_aa_ww}
\end{figure}

In this section, we study an EW process
\begin{align}
  \gamma + \gamma \to W^- + W^+. 
  \label{proc:aaWW}
\end{align}
There are three Feynman diagrams for the process,
as shown in Figs.\,\ref{fig:diagram_aa_ww}(a), (b), (c), corresponding to
$t$- and $u$-channel $W$ exchange, and the contact
term, respectively.
We study the cross section for the process with both
$W$ bosons decay into massless fermion pair, in order to avoid introducing
the weak boson polarization vectors in general LC gauge,
which are not helicity eigenstates. 
Specifically, we examine leptonic decays, 
\begin{eqnarray}
W^- \to e^- {\bar\nu}_e,\quad W^+ \to \mu^+ \nu_\mu.
\end{eqnarray}
Once $W^\pm$ decays, we will have three weak boson
propagators in the diagrams (a) and (b), and two
weak boson propagators in the diagram (c).

In the LC gauge, we assign the same LC gauge vector
$n^\mu$ for all the $W^\pm$ propagators.
In this report, we examine the following three gauge
vectors:
\begin{subequations}
\begin{align}
n^\mu &= (1, 0, 0, 1) 
\label{subeq:n1001}  \\
n^\mu &= (1, 1, 0, 0) 
\label{subeq:n1100}  \\
n^\mu &= (1, \overrightarrow{p}_{W^-}^{}/|\overrightarrow{p}_{W^-}^{}|)  
\label{subeq:n1pw}
\end{align}
\end{subequations}
The first choice\,\eqref{subeq:n1001}, is the same as eq.\,\eqref{eq:LCvector} employed in previous QED and QCD studies.
The second choice\,\eqref{subeq:n1100} specifies the scattering
plane of the process, which takes place in the
$(x,z)$ plane.
The third gauge vector\,\eqref{subeq:n1pw} is chosen along
the $W^-$ momentum direction, opposite to the $W^+$ momentum direction.

\begin{figure*}[t]
\center
\includegraphics[width=1\textwidth]{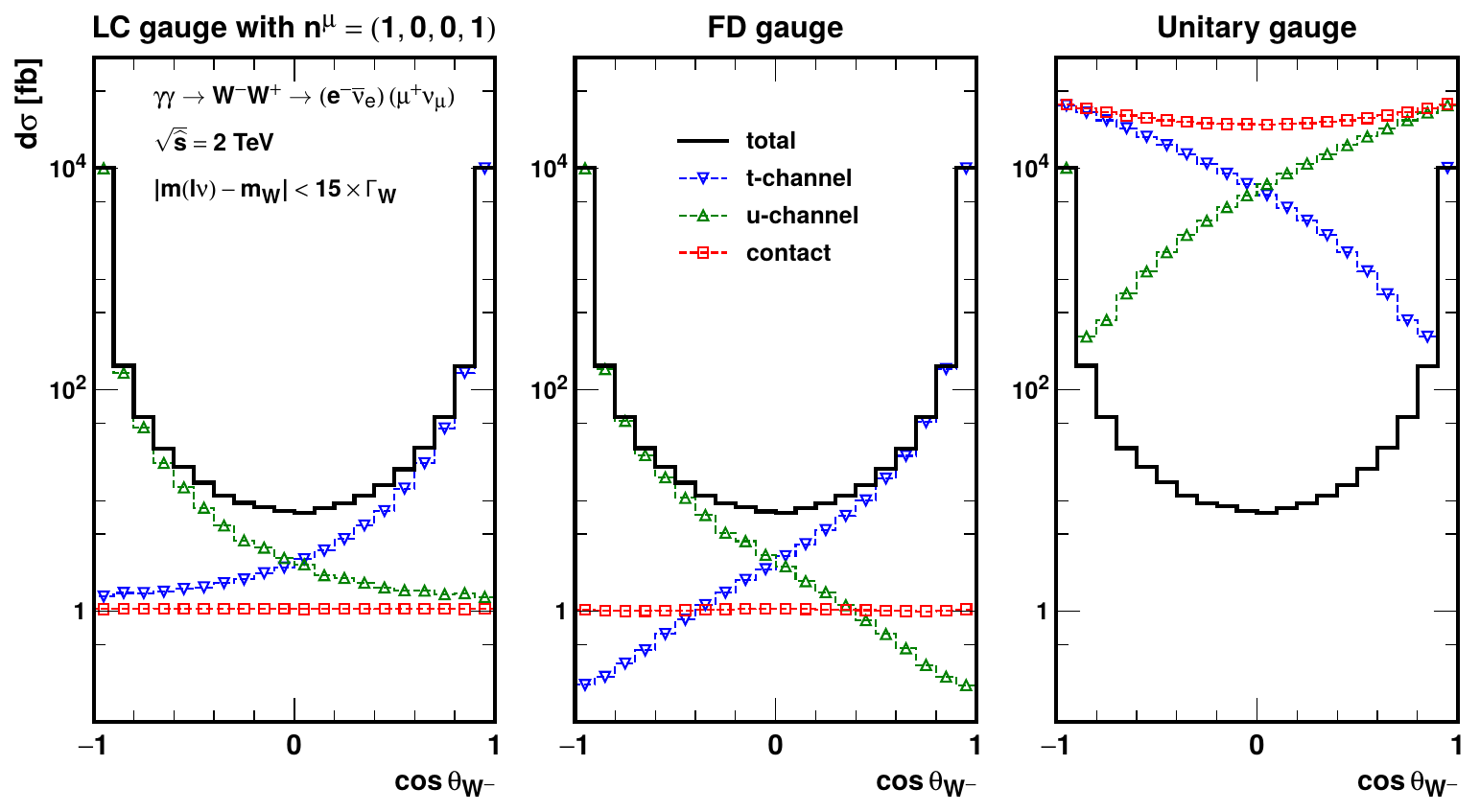}
\hspace*{1cm}(a)\hspace*{5.5cm}(b)\hspace*{5.5cm}(c)
\caption{Distribution of the scattering angle for $\gamma\gamma\to W^-W^+$ at $\sqrt{s}=2$~TeV 
in three different gauges:
(a) LC gauge with $n^\mu=(1,0,0,1)$, (b) FD gauge, and (c) the Unitary gauge. 
The invariant masses of the final lepton pairs are integrated out in the $W$ resonant region,
$|m(e^- \bar{\nu}_e)-m_W|, |m(\mu^+ \nu_\mu)-m_W| < 15~{\rm \Gamma}_W$.}
\label{fig:aa_ww}
\end{figure*}

In Fig.\,\ref{fig:aa_ww}, we show the differential cross section
with respect to $\cos\theta_{W^-}$ in the $\gamma\gamma$ collision
rest frame.
Here $\theta_{W^-}$ is
the scattering angle of $W^-$, or that of
$e^-$ and $\bar{\nu}_e$ pair, measured from one of the
incoming photon momentum direction.
The energy is set high at $\sqrt{s} = 2$~TeV, in order
to demonstrate that the LC gauge amplitudes for
the weak boson scattering processes are free from
the terms which grow with the weak boson energy,
as observed already in 1987 by Kunszt and Soper\!\cite{Kunszt:1987tk},
by Dams and Kleiss in 2004~\cite{Dams:2004vi}
and more recently for the process
$\gamma\gamma \to W^+ W^-$ by Bailey and Harland-Lang\!\cite{Bailey:2022wqy}
\footnote{
The authors of refs.\!\cite{Kunszt:1987tk,Dams:2004vi,Bailey:2022wqy} adopt general axial gauge,
while we can regard our LC gauge as a special case of
the axial gauge.
}

%

The cross section is obtained by integrating out the invariant masses of the final lepton pairs in the range 
\begin{eqnarray}
\left|m\left(e^-\bar{\nu}_e\right)-m_W\right|,
\left|m\left(\mu^+{\nu}_\mu\right)-m_W\right|<15~{\rm \Gamma}_W
\end{eqnarray}
for $m_W=80.4$ GeV and ${\rm \Gamma}_W=2.1$ GeV\!\cite{ParticleDataGroup:2022pth}. Numerical results are obtained for the EW couplings $\alpha=\alpha(m_Z)=1/128$ and $\sin^2\theta_{W}=0.231$.

The Unitary gauge results as displayed in Fig.\,\ref{fig:aa_ww}(c)
show clearly the $(E_W/m_W)^2$ growth of the
individual Feynman amplitude, where the amplitude
with the contact interaction diagram, Fig.\,\ref{fig:diagram_aa_ww}(c), illustrated by the red dashed curve dominates at all
$\cos\theta_{W^-}$.
All three amplitudes are large at around 
$\cos\theta_{W^-}=0$, where the cancellation
makes the physical cross section more than three
orders of magnitude smaller than the square of
the individual amplitudes.
This has been well known, since even the conserved
current of massless lepton pair has longitudinal
polarization component, which grows with energy
in the Unitary gauge.
In fact, when massless lepton currents are attached to
the weak bosons, whether in the initial state or in the final state of the weak boson scattering processes,
there is no gauge dependence in
the weak boson propogators within the covariant $R_\xi$ gauge,
since the current is conserved (no terms with
$q^\mu q^\nu$ survive) and it has no Goldstone boson
coupling in the massless limit.
The huge growth of the individual amplitude at high energies, as shown in Fig.\,\ref{fig:aa_ww}(c), is common to all the covariant $R_\xi$ gauges.

On the contrary, in Figs.\,\ref{fig:aa_ww}(a) and (b),
there is no subtle cancellation among the  three
amplitudes in the LC gauge with the gauge vector\,\eqref{subeq:n1001}
and in the FD gauge.
We note that in both gauges, single diagram dominates
the full amplitude at $|\cos\theta_{W^-}| \approx 1$.
In particular, the $t$-channel exchange amplitude
Fig.\,\ref{fig:diagram_aa_ww}(a) dominates the forward region
$\cos\theta_{W^-} \gtrsim 0.6$,
while the $u$-channel exchange amplitude dominates
the backward region
$\cos\theta_{W^-} \lesssim-0.6$.
This has been expected for the FD gauge, as it is
designed to make individual Feynman amplitude
behave as products of the Feynman propagator factors
connected by the physical splitting amplitude at
each vertex.
The LC gauge results with the gauge vector\,\eqref{subeq:n1001}
are very similar to the FD gauge. It can be understood as follows.
In the limit of $|\cos\theta_{W^-}| = 1$, the
$t$- and $u$-channel exchange weak boson three
momenta are either along the positive or the negative
$z$-axis.
In this limit, where the cross section is the largest,
the LC gauge vector agrees with the FD gauge vector,
up to the sign of the three-vector piece.
This sign dependence does not affect the amplitudes in the process\,\eqref{proc:aaWW}
because the amplitude is symmetric in the exchange
of two incoming photons.

On the other hand, the Unitary gauge results show that even
near the $|\cos\theta_{W^-}|=1$ limits, the distribution behavior is unusual.
The $t$-channel exchange amplitude, whose absolute
square is given by the blue curve in Fig.\,\ref{fig:aa_ww}(c),
{\it decreases} with $\cos\theta_{W^-}$,
quite opposite of what we naively expect from the
Feynman's propagator behavior.
Then at the last $\cos\theta_{W^-}$ bin of the
histogram, $0.95 < \cos\theta_{W^-} <1$, the
$t$-channel exchange amplitude dominates the
physical amplitude, and the other two amplitudes
with larger magnitudes cancel out almost exactly.
Although we do not look into more details on this,
it is clear that the Unitary gauge, or any other
covariant $R_\xi$ gauge amplitudes are often impractical
in obtaining physical insights out of individual scattering
amplitudes.

Next, we study the LC gauge results with
other gauge vectors, eqs.\,\eqref{subeq:n1100} and\,\eqref{subeq:n1pw}.
We find unexpectedly that our Monte Carlo
integration program doesn't produce results for these
two gauge vectors.
Although the problem associated with the LC gauge
singularity may have been well known\!\cite{Leibbrandt:1987qv}, we would like to share in this report what we learn by using our new automatic
amplitude calculation code.

\begin{figure*}[t]
\center
\includegraphics[width=1\textwidth]{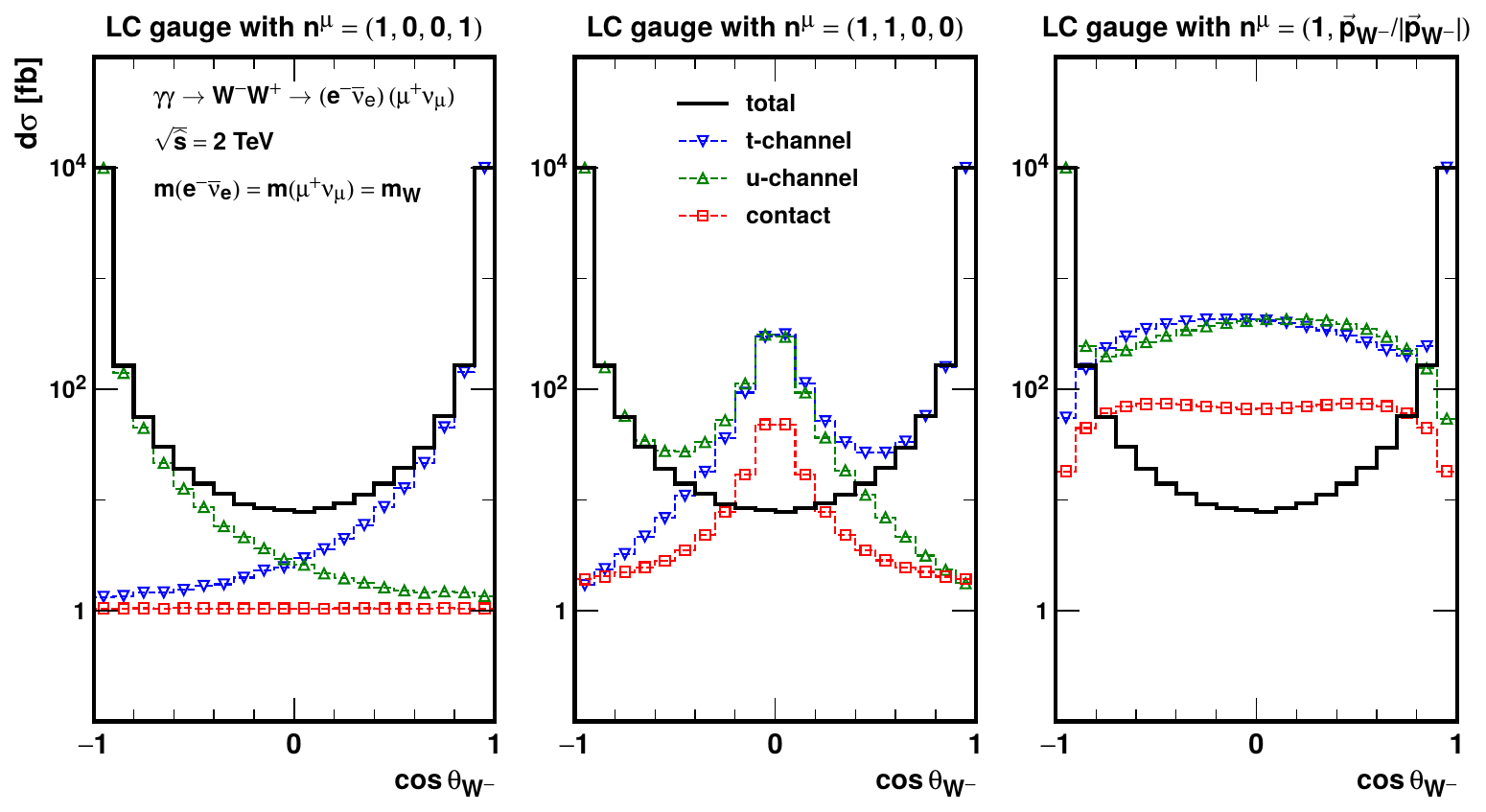}
\hspace*{1cm}(a)\hspace*{5.5cm}(b)\hspace*{5.5cm}(c)
\caption{Distribution of the scattering angle for $\gamma\gamma\to W^-W^+$ at $\sqrt{s}=2$~TeV 
in the LC gauge with different LC vectors, when both $W^+$ and $W^-$ are exactly on the mass shell.}
\label{fig:aa_ww_LC}
\end{figure*}

We first note that all the LC gauge amplitudes give
exactly the same cross section when both $W$'s are
on the mass-shell:
\begin{eqnarray}
m(e^- \bar{\nu}_e) = m(\mu^+ \nu_\mu) = m_W.  
\label{eq:mw}
\end{eqnarray}
The three plots of Fig.\,\ref{fig:aa_ww_LC} show the differential
cross section with respect to $\cos\theta_{W^-}$
when the final state is constrained to satisfy the
double on-shell conditions\,\eqref{eq:mw}.
The black solid histograms are obtained from the physical amplitudes where the three Feynman diagrams of Fig.\,\ref{fig:diagram_aa_ww} are summed before squaring.
The gauge invariance of the physical amplitudes are clearly observed from the exact agreement of the three black histograms in Fig.\,\ref{fig:aa_ww_LC}.
The individual Feynman diagram contributions,
however, are very different among the three
LC gauges.
The gauge vector\,\eqref{subeq:n1001} gives distributions which are
consistent with the behavior of the corresponding
Feynman propagator in the $t$- and $u$-channels,
and its absence in the contact term, just as
in the FD gauge.
For the gauge vector\,\eqref{subeq:n1100}, each Feynman amplitude gives
approximately the same magnitude at
$\left|\cos\theta_{W^-}\right|\gtrsim0.6$ as that of the LC gauge\,\eqref{subeq:n1001}.
All the three Feynman amplitudes, however, have large magnitude
at around $\cos\theta_{W^-}\approx 0$, where
destructive interference among the three amplitudes
give the physical cross section.
In case of the gauge vector\,\eqref{subeq:n1pw}, the enhancement
of the magnitude of all three individual Feynman
amplitudes is observed in the wide region of the
scattering angle besides near the two edges around
$\left|\cos\theta_{W^-}\right|\approx 1$.
We study in detail the LC gauge vector dependence of
each amplitude in order to understand their behaviors
shown in these figures.

\begin{figure*}[t]
  \center
\includegraphics[width=2\columnwidth]{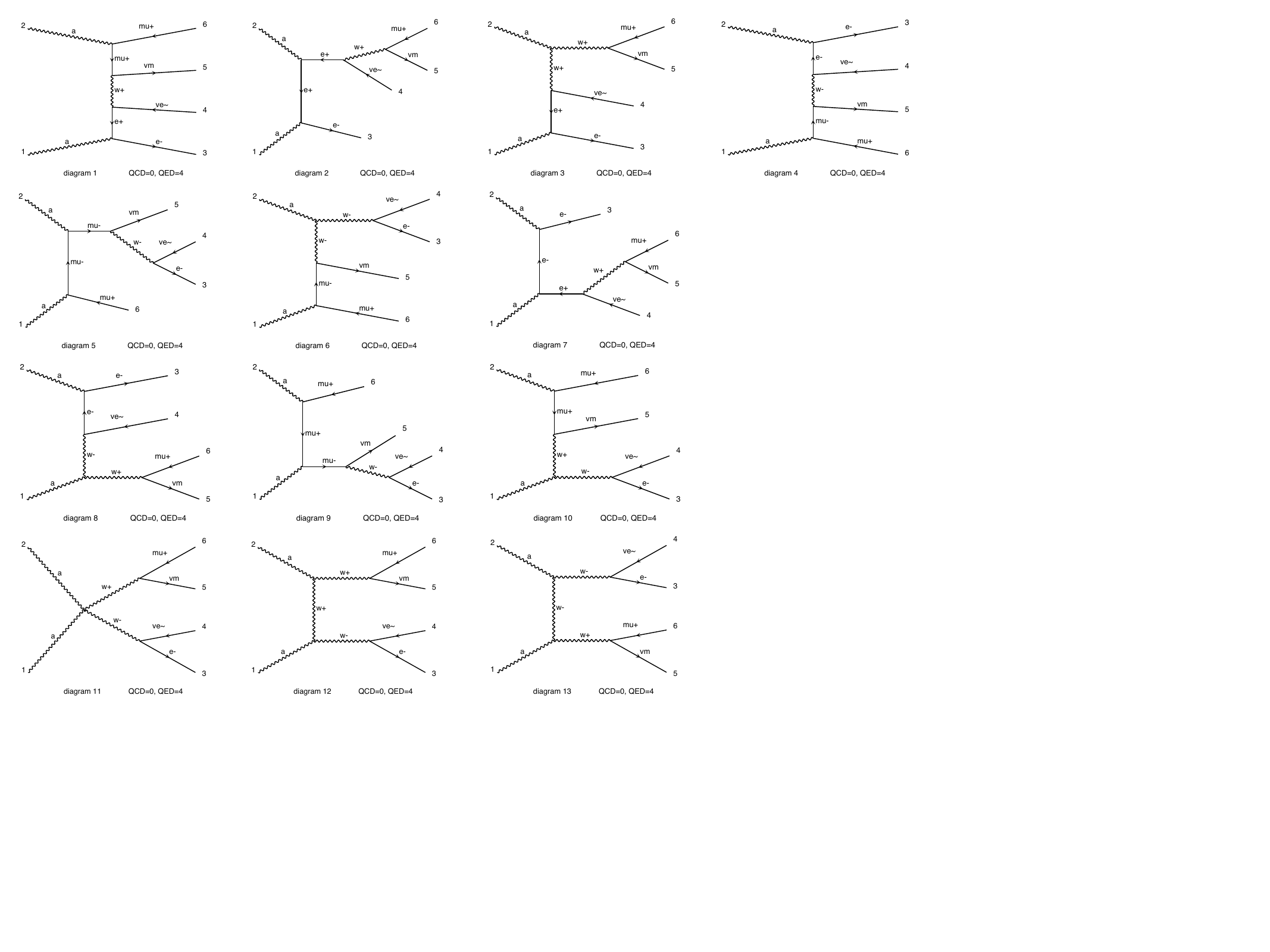}
\caption{Feynman diagrams for $\gamma\gamma\to e^-\bar{\nu}_e\mu^+\nu_\mu$, generated by {\tt Madgraph}.}
\label{fig:Feyn13_aaWW}
\end{figure*}

We first note that the enhancement found for the
contact interaction diagram, Fig.\,\ref{fig:diagram_aa_ww}(c), which is
given by red curves in Fig.\,\ref{fig:aa_ww_LC}, can only come
from the two weak boson propagators attached to
the final lepton pairs.
The lepton current,
\begin{eqnarray}
J^\mu = u_L(p_e)^\dagger \sigma_-^\mu v_L(p_{\bar\nu}) 
\end{eqnarray}
can be expressed in general as a summation over three
$W$ boson helicity components as follows:
\begin{eqnarray}
J^\mu
  &=&
\left(g^{\mu}_{}{}_{\nu} -\frac{q^\mu q_\nu}{q^2}\right) J^\nu
\nonumber\\
  &=&
\left(-\sum_{h=\pm 1,0} \eps^\mu(q,h)^*~~ \eps_\nu(q,h)\right)J^\nu
\nonumber\\
  &=&
-\sum_{h=\pm 1,0} \eps^\mu(q,h)^* ~~ \eps(q,h)\cdot J,
  \end{eqnarray}
where we assume zero lepton masses so that the
current is conserved and has no Goldstone boson
coupling.
In the above expression, the term
\begin{eqnarray}
\eps(q,h)\cdot J 
\end{eqnarray}
gives Wigner's $d$-functions~\cite{Wigner:1939cj} in the rest frame of
the decaying weak boson, or the splitting amplitudes\!\cite{Hagiwara:2009wt}
in boosted frames.
Since the transverse polarization components $h=\pm 1$
do not grow with the weak boson energy, it is only
the $h=0$ component which grows with energy.
In the Unitary gauge, or in an arbitrary covariant
$R_\xi$ gauge, it is this $h=0$ component which grows with the weak boson energy.
In the FD gauge, this $h=0$ component is replaced by
the $\tilde{\epsilon}^\mu(q,0)$ term, whose magnitude decreases
with energy.

Let us examine what happens to the $h=0$ component of
the leptonic current in the general LC gauge:
\begin{eqnarray}
P_{\rm LC}{}^\mu{}_\nu \epsilon^\nu(q,0)
  =
P_{\rm LC}{}^\mu{}_\nu \left(\frac{q^\mu}{Q}+\tilde{\epsilon}^\nu(q,0)\right).
\end{eqnarray}
For our example of 
$E_W=1~{\rm TeV}$,
 the magnitude of
$\tilde{\epsilon}^\nu(q,0)$ decreases by a factor of $e^{-\eta}\approx0.04$ for
$\cosh\eta=E_W/m_W$,
and we can safely neglect its contribution against 
$q^\mu/Q\sim{\cal O}(E_W/m_W)$.
Therefore, it is only the behavior of the scalar current term
$q^\mu/Q$ we should examine.

We first note that the general LC gauge polarization tensor can be decomposed as
\begin{eqnarray}
P_{\rm LC}^{}{}^{\mu\nu} = P_T^{\mu\nu}
  +
q^2 \frac{n^\mu  n^\nu}{(n \cdot q)^2}  ,
\end{eqnarray}
where the so-called `transverse' polarization tensor
satisfies
\begin{eqnarray}
P_T^{\mu\nu} q_\mu = P_T^{\mu\nu} n_\nu = 0. 
\label{eq:transPolTensor}
\end{eqnarray}
Because of the property\,\eqref{eq:transPolTensor}, the scalar current
component of the leptonic current reduces to
\begin{eqnarray}
P_{\rm LC}{}^\mu_{\ \nu}~ \frac{q^\nu}{Q}
  =
Q ~\frac{n^\mu}{n \cdot q} . 
\label{eq:PLCmunu}
\end{eqnarray}
Since $Q \approx m_W$ around the resonance peak,
and since $n^\mu$ is a constant gauge vector,
it is the gauge term $n\cdot q$ in the denominator which we should examine.
%
\begin{figure*}[t]
\center
\includegraphics[width=0.8\textwidth]{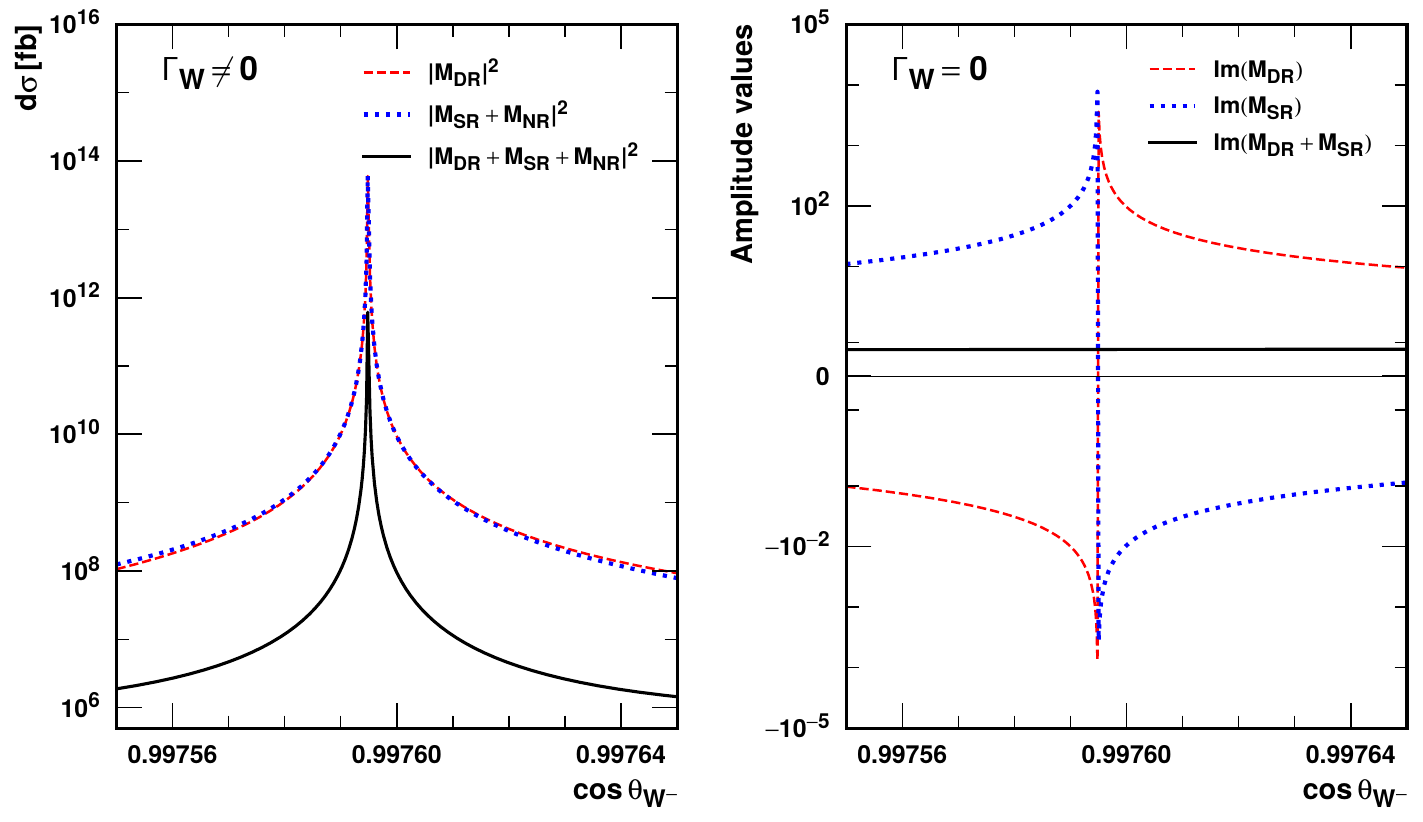}
\hspace*{0.5cm}(a)\hspace*{6.8cm}(b)
\caption{
(a) Differential cross section $d\sigma/d\cos\theta_{W^-}$ for the process $\gamma\gamma\to e^-\bar{\nu}_e\mu^+\nu_\mu$ at $\sqrt{ s}=2$~TeV in the LC gauge with $n^\mu=(1,\sin\theta_{W^-},0,\cos\theta_{W^-})= (1, \protect\vect{p}_{W^-}/|\protect\vect{p}_{W^-}|)$, where $\theta_{W^-}$ is the polar angle of the $e^-\bar\nu_e$ pair. 
The invariant masses of the lepton pair are fixed at $m(\mu^+\nu_\mu)=69.3$~GeV and $m(e^-\bar\nu_e)=m_W$.
The red dashed curve shows the contribution of the three double resonant diagrams (${\cal M}_{\rm DR}$), while the blue dotted curve shows the contribution of all the other diagrams, single resonant (${\cal M}_{\rm SR}$) or non-resonant (${\cal M}_{\rm NR}$).
The black solid curve shows the sum of all 13 Feynman diagrams.
(b)
Imaginary part of the amplitudes when the photon helicities are $\lambda_1=\lambda_2=-1$ and $e^-$ and $\mu^+$ momenta are along the $W^+(\mu^+\nu_\mu)$ momentum direction in the $\gamma\gamma$ rest frame.
We set ${\rm \Gamma}_W=0$ in the off-shell $W^-$ propagator.
}
\label{fig:aa_ww_decay}
\end{figure*}

Let us write down the four momentum of $W^\pm$
in the colliding photon rest frame.
We find
\begin{subequations}
\begin{align}
p^\mu_{W^-}
  &=
(E_{W^-},p^*\sin\theta_{W^-},0,p^*\cos\theta_{W^-}) 
\\
p^\mu_{W^+}
 & =
(E_{W^+},-p^*\sin\theta_{W^-},0,-p^*\cos\theta_{W^-}) 
\end{align}
\end{subequations}
where
\begin{eqnarray}
&&E_{W^\pm} = \frac{\sqrt{s}}{2}
(1+\frac{p_{W^\pm}^2-p_{W^\mp}^2}{s}) 
\end{eqnarray}
 and
\begin{eqnarray}
p^* = \frac{\sqrt{s}}{2}
{\bar\beta}(\frac{p_{W^-}^2}{s},\frac{p_{W^+}^2}{s})  
\end{eqnarray}
with 
\begin{eqnarray}
\bar{\beta}(a,b)=\sqrt{1-2(a+b)+(a-b)^2}
\end{eqnarray}
is the common momentum of the $W$ pair in the c.m. frame.

For the gauge vector\,\eqref{subeq:n1001}, the product is
\begin{eqnarray}
n\cdot p_{W^\pm} = E_{W^\pm} \pm p^*\cos\theta_{W^-}
\end{eqnarray}
which can become small around $\cos\theta_{W^-}=\pm 1$,
but after contracting all the four currents in the contact-term amplitudes Fig.10(c), the
gauge terms proportional to $n^\mu$ cancel out:
They vanish in the product of the $W^-$ and $W^+$ currents
because of $n\cdot n=0$, whereas the product of $W^\pm$
and the initial photon polarization vectors vanish,
since
\begin{eqnarray}
n \cdot \epsilon(p_\gamma,\pm 1) = 0
\end{eqnarray}
for both photons with momenta along the z-axis.

In case of the gauge vector\,\eqref{subeq:n1100}, we find
\begin{eqnarray}
n\cdot p_{W^\pm} = E_{W^\pm} \pm p^*\sin\theta_{W^-}
\end{eqnarray}
and one of which can be small at
$\cos\theta_{W^-}\approx 0$.
In the contact interaction amplitude, the enhanced
current survives when contracted with the initial
photon polarization vectors, because
\begin{eqnarray}
\left|n \cdot \epsilon(p_\gamma,\pm 1)\right|
  =
\left|-\epsilon(p_\gamma,\pm 1)^1\right| = \frac{1}{2} 
\end{eqnarray}
for both photons. 
The enhancement in the red histogram around 
$\cos\theta_{W^-}\approx 0$ observed in Fig.\,\ref{fig:aa_ww_LC}(b)
is hence due to the LC gauge term.

Finally, for the gauge vector of\,\eqref{subeq:n1pw}, we find
\begin{eqnarray}
n\cdot p_{W^\pm} = E_{W^\pm} \pm p^* 
\label{eq:npw}
\end{eqnarray}
which is independent of $\cos\theta_{W^-}$.
In fact, the above term grows with energy for the
$W^+$ propagator, because the three-vector direction
$\vect{n} = \vect{p}_{W^-}/|\vect{p}_{W^-}|$ is opposite of
the $W^+$ momentum direction in the c.m.\,frame,
which is the prescription for the FD gauge.
It is hence only the $W^-$ propagator whose gauge
term can grow with energy.
Because the enhancement factor in the $W^-$ propagator
is independent of the scattering angle, we observe
broad enhancement in Fig.\,\ref{fig:aa_ww_LC}(c).
The enhancement effect diminishes near
$\cos\theta_{W^-} = \pm 1$, because the gauge vector
becomes orthogonal to one of the incoming photon
polarization vectors in the limits.

Although we understand the gauge artefacts
observed for the contact interaction term as above, we find
that the most serious problem arises in the $t$- and $u$-channel exchange amplitudes, from their gauge terms.
The four momentum of the weak boson exchanged in the
$t$-channel can be expressed as follows:
\begin{eqnarray}
q^\mu
  &=&
p^\mu(\gamma_1) -p^\mu(W^-)
\nonumber\\
&  =&
(E-E_{W^-}, -p^*\sin\theta_{W^-}, 0, E-p^*\cos\theta_{W^-})
\label{eq:t-4mom}
\end{eqnarray}
with $E=\sqrt{s}/2$.
Let us show the gauge term for our three gauge vectors.
For\,\eqref{subeq:n1001}, we find
\begin{eqnarray}
n\cdot q
=
-( E_{W^-} -p^*\cos\theta_{W^-}).
\end{eqnarray}
This term can never vanish as long as both $W^\pm$
invariant mass is in the resonance region.
For\,\eqref{subeq:n1100}, we find
\begin{eqnarray}
n\cdot q
  =
E-E_{W^-} +p^*\sin\theta_{W^-}.
\end{eqnarray}
When both $W$'s are on-shell,  $E_{W^-}=E$,
and hence the gauge term vanishes at
$\sin\theta_{W^-}=0$, or at
$\cos\theta_{W^-}=\pm1$. 
Once the invariant mass of $W^-$ is larger than that
of $W^+$, the zero appears in the physical region.
Although the LC gauge pole singularity can be
integrated out in loop calculation e.g.\ by taking
the principal value integral\!\cite{Leibbrandt:1987qv}, once the
pole appears in the physical region of the
phase space integral, the MC integral fails
instantly, since the singular terms appear in
the absolute value square of the amplitudes.

Let us study how the LC gauge pole singularity
appears in the resonant amplitudes, and how they
are cancelled in the physical amplitudes, by
using the gauge vector\,\eqref{subeq:n1pw}.
The four momentum of the $W$ boson exchanged in the
$t$-channel is given by eq.\,\eqref{eq:t-4mom}
and the gauge term for\,\eqref{subeq:n1pw} becomes
\begin{eqnarray}
n\cdot q
  =
E-E_{W^-}+p^*-E*\cos\theta_{W^-}.
\end{eqnarray}
The $n\cdot q=0$ pole can appear at
\begin{eqnarray}
\cos\theta_{W^-} = (E-E_{W^-}+p^*)/E .
\label{eq:thetaW}
\end{eqnarray}
When both $W$'s are on-shell, this happens at
\begin{eqnarray}
\cos\theta_{W^-} = \sqrt{1-(m_W/E)^2},
\end{eqnarray}
which is 0.9968 at $\sqrt{s}=2$~TeV.
We find, however, that this pole does not show up
in the amplitudes because each on-shell amplitude
vanishes exactly at the corresponding pole.

We find that the above vanishing of individual
Feynman amplitude at the pole of the LC gauge term
does not hold when one or both of the weak bosons
is off-shell. 
There appear non-zero amplitudes which are proportional
to the off-shellness of decaying $W$'s, 
$\left(m(e^- \bar{\nu}_e\right)^2 -m_W^2)$ and/or 
$\left(m(\mu^+ \nu_{\mu}\right)^2 -m_W^2)$,
as residues of the pole term.
We further find that the singular terms do not
cancel even after summing up the three resonant
diagrams of Figs.\,\ref{fig:diagram_aa_ww}(a),(b),(c).

We therefore evaluate all the 13 diagrams contributing
to the process
\begin{eqnarray}
\gamma \gamma \to e^- {\bar\nu}_e \mu^+ \nu_\mu
\end{eqnarray}
as shown in Fig.\,\ref{fig:Feyn13_aaWW}, generated by {\tt MadGraph}\!\cite{Stelzer:1994ta,Alwall:2011uj,Alwall:2014hca}.
We note that the three double resonant (DR) diagrams of Fig.\,\ref{fig:diagram_aa_ww} are now labeled as 11, 12, 13 at the bottom of Fig.\,\ref{fig:Feyn13_aaWW}. 
The diagrams 2, 3, 7, 8 have $W^+$ propagator, while the diagrams 5, 6, 9, 10 have $W^-$ propagator only, and we call them single resonant (SR) diagrams. The diagrams 1 and 4 have no $W^{\pm}$ propagator, and are labeled as non-resonant (NR).

In Fig.\,\ref{fig:aa_ww_decay}(a), we show the
$\cos\theta_{W^-}$ distribution when
\begin{subequations}
\begin{eqnarray}
m(e^- {\bar\nu}_e)&=&69.3~{\rm GeV}\approx m_W-5.3~{\rm \Gamma}_W ,
\\
m(\mu^+ \nu_\mu)&=&m_W=80.4~{\rm GeV},
\label{eq:mw804}
\end{eqnarray}
\label{eq:mw693804}
\end{subequations}
 at $\sqrt{s}=2$~TeV.
The red dashed curve shows the contribution of
the sum of the three double resonant amplitudes,
\begin{eqnarray}
\left|{\cal M}_{\rm DR}\right|^2=\left|\sum_{k=11}^{13} {\cal M}_k\right|^2.  
\end{eqnarray}
Shown by the blue dashed curve is for the sum of all the other amplitudes
\begin{eqnarray}
\left|{\cal M}_{\rm SR}+{\cal M}_{\rm NR}\right|^2
=
\left|\sum_{k=1}^{10} {\cal M}_k\right|^2,
\end{eqnarray}
which sum over the 8 single resonant and 2 non-resonant amplitudes.
We can tell from the red and blue dashed curves that they
have the same singular behavior at
$\cos\theta_{W^-}$\\$\approx 0.9976$,
which is the pole position of eq.\,\eqref{eq:thetaW} for the
invariant masses of eq.\,\eqref{eq:mw693804}.
Although the red and blue dashed curves in Fig.\,\ref{fig:aa_ww_decay}(a) confirm
that the sum of the three double-resonant amplitudes (${\cal M}_{\rm DR}$) and that of
the single and non-resonant amplitudes (${\cal M}_{\rm SR}+{\cal M}_{\rm NR}$) have the same singular
behavior, we find that the total sum of all the 13 diagrams in Fig.\,\ref{fig:Feyn13_aaWW}, 
\begin{eqnarray}
\left|{\cal M}_{\rm DR}+{\cal M}_{\rm SR}+{\cal M}_{\rm NR}\right|^2
=
\left|\sum_{k=1}^{13} {\cal M}_k\right|^2
\end{eqnarray}
still gives the singular behavior,
as shown by the black solid curve in the same plot.
The magnitude of the singular term is significantly
reduced, and hence there is a hint of cancellation among the 13 amplitudes of Fig.\ref{fig:Feyn13_aaWW}\,.

We find that this non-cancellation of the LC gauge pole singularity is due to the width term in the $W$ boson propagator, unitarized by the Breit-Wigner prescription~\cite{Breit:1936zzb}. 
At the kinematical point on top of the pole,
$n\cdot p_W=0$,
the non-vanishing off-shell amplitudes are proportional to 
$p_W^2-m_W^2$.
This term cancels the $W$ propagator factor exactly, only if the width is set to be zero. 
In the presence of the finite width term, the cancellation is not exact, and the pole singularity in the resonant amplitudes survive even after summing over all the non-resonant amplitudes. 

In order to demonstrate cancellation of the LC gauge pole singularity between the resonant and non-resonant amplitudes, we show in Fig.14(b) the amplitude of the double resonant diagrams, and that of the single $W^+$ resonant  diagrams for the lepton-pair invariant masses of eq.(76). 
We show only the amplitude when both incoming photons are left-handed 
$\left(\lambda_{\gamma_1}=\lambda_{\gamma_2}=-1\right)$,
and when the final $e^-$ and $\mu^+$ momenta are along the $W^+$ $(\mu^+\nu_\mu)$ momentum direction. 
With this set up the double resonant amplitude is proportional to the helicity amplitude for 
\begin{eqnarray}
\gamma_1\left(-1\right)+\gamma_2\left(-1\right)
\to
W^+\left(+1\right)+
W^-\left(+1\right),
\end{eqnarray}
where $\pm1$ inside parentheses denote helicities.
We keep the width factor of the $W^+$ propagator so that the amplitude is finite and pure imaginary on top of the $W^+$ boson mass shell,\,\eqref{eq:mw804}. 
We set the width of off-shell $W^-$ propagator to zero, in order to demonstrate the cancellation of the LC gauge pole singularities between the double resonant amplitudes (${\cal M}_{\rm DR}={\cal M}_{11}+{\cal M}_{12}+{\cal M}_{13}$) and the single $W^+$ resonant amplitudes (${\cal M}_{\rm SR}={\cal M}_2+{\cal M}_3+{\cal M}_7+{\cal M}_8$).
These amplitudes are pure imaginary (due to on-shell $W^+$), and we show the imaginary part of $M_{\rm DR}$ by the red-dashed line, that of ${\cal M}_{\rm SR}$ by the blue-dashed line, and their sum by the black line. 
We can clearly observe exact cancellation of the LC gauge pole singularity between the ${\cal M}_{\rm DR}$ and the ${\cal M}_{\rm SR}$ amplitudes. 
Similar cancellation takes place in the real part of the amplitudes between the $W^-$ SR amplitudes and NR amplitudes, with significantly smaller magnitudes.

The apparent LC gauge vector dependence of the amplitudes is hence the artefact of the Breit-Wigner unitarization\,\cite{Breit:1936zzb} of the $W$ boson propagator, which violates the order-by-order gauge invariance of the perturbative amplitudes.

Since it is not practical to sum up all the
non-resonant amplitudes before unitarizing
the resonant $W$ amplitudes, we propose that
the LC gauge vector should be carefully
chosen such that the LC gauge term does not
have a pole in the physical region of all the 
contributing gauge boson propagators.
Among our examples, the choice\,\eqref{subeq:n1001} should 
be safe, as long as the final state particles are
massive or has large $p_T$.

\section{Summary}\label{sec:summary}
In this report, we show that the FD gauge propagators introduced in refs.\!\cite{Hagiwara:2020tbx,Chen:2022gxv} can be obtained from the LC gauge propagators, or the Green's functions of
the equation of motion (EOM) of the free gauge fields in the LC gauge,
not only in QED and QCD, but also in the EW theory
with massive weak bosons.
In particular, the $5 \times 5$ representation of
the weak boson propagators is directly obtained from
the EOM of the free weak boson and its associate
Goldstone boson fields in the LC gauge.
The FD gauge propagators are then obtained simply
by choosing the LC gauge vector along the opposite direction of the three-momentum of the off-shell gauge boson.

Because the general LC gauge propagators and
the scattering amplitudes have identical form as those in the FD gauge, we modify the new {\tt HELAS}\!
codes\cite{Hagiwara:2020tbx,Chen:2022gxv} to allow an arbitrary LC gauge vector
which is common among all the propagators in the
amplitudes. 
With this minimum modification of the programs
developed in refs.\!\cite{Hagiwara:2020tbx} and~\cite{Chen:2022gxv}, we are able to
obtain helicity amplitudes of all the SM processes
in an arbitrary LC gauge.

Since helicity amplitudes for all $2 \to 2$ processes
are gauge invariant in QED and QCD, we have studied 
a few $2\to 3$ processes, $\ell\bar{\ell}^\prime\to \ell\bar{\ell}^\prime\gamma$, $\ell{\ell}^\prime\to \ell{\ell}^\prime\gamma$ and
$gg\to ggg$, in section~\ref{sec:results}.
We find that LC gauge amplitudes don't
improve cancellation among diagrams in general, because only when the LC gauge vector orientation is
along or approximately along the FD gauge vector of a particular current, the off-shell current
components proportional to its four momentum are suppressed.
We note that a specific choice of the LC gauge vector gives soft gluon radiation patterns efficiently, confirming the observation in parton shower studies~\cite{Nagy:2007ty,Nagy:2014mqa}.

In case of the EW theory amplitudes, $2\to  2$
processes like $\gamma\gamma \to W^+ W^-$ have
gauge dependence, and we study the amplitudes in detail in section~\ref{sec:EW}.
We confirm that, just like
in the FD gauge, the LC gauge amplitudes are free
from subtle cancellation among interfering amplitudes~\cite{Bailey:2022wqy} 
which are unavoidable in all covariant gauges,
including $R_\xi$ and Unitary gauges.
We also find that we should choose the LC gauge vector
cautiously, such that no pole due to the vanishing gauge term, $n\cdot q=0$, should
appear in the physical region of any of the gauge
boson propagators.

From the perspective of high energy computation tools development, our LC gauge quantization approach
has an absolute advantage over the previously proposed
method~\cite{Chen:2022gxv} of modifying the amplitudes generated
in the Unitary or $R_\xi$ gauges.
If we start from the Unitary gauge, as in ref.\!\cite{Chen:2022gxv}, although the number of Feynman diagrams
can be reduced due to the absence of the Goldstone
boson propagation, some of the fundamental vertices
like the four-point neutral Goldstone boson coupling
are absent from the Feynman rule.
As a consequence, the authors of ref.\!\cite{Chen:2022gxv} added a few
vertices to the SM interactions in order to generate 
appropriate Feynman diagrams and the corresponding amplitudes. 
If we start from the $R_\xi$ gauge, things become even more complicated.
 It is because all the Goldstone boson
propagators should first be cancelled against the
corresponding $R_\xi$ gauge boson propagators which
share the same four momentum, by using the BRST identities. 
At this stage, all the amplitudes with the Goldstone boson propagators are removed, and the vertices which are missing in the Unitary gauge should be re-introduced. 
Although there are only four such
vertices in the SM of the EW theory~\cite{Chen:2022gxv}, we anticipate more missing vertices, in models beyond the SM
or even in the SM effective field theory (SMEFT) .

On the other hand, in the LC gauge quantization of the
weak bosons, no unphysical particles propagates because all the Goldstone bosons are the \nth{5} component
of the corresponding physical weak bosons. 
The Feynman rules are then straightforward,
since we do not distinguish the first 4 and the \nth{5}
components of the weak bosons when generating Feynman
diagrams.
All the vertices are straightforward to obtain
once the Higgs sector of the model is specified.
The method can be applied for all SMEFT operators,
and for all BSM models with spontaneously broken gauge symmetries.
We therefore believe that the next generation of
scattering amplitude generators should adopt the
LC gauge form of the massless and massive gauge
boson propagators.
The FD gauge amplitudes are then obtained by choosing the gauge vector along the opposite direction of each off-shell gauge boson momentum.

\section*{Acknowledgements}
JK and YZ would like to thank KEK Theory Center for hospitality where part of the work is done.
JMC is supported by Fundamental Research Funds for the Central Universities of China NO. 11620330.
The work was supported in part 
by World Premier International Research Center Initiative (WPI Initiative), MEXT, Japan, and also
by JSPS KAKENHI Grant No. 20H05239, 21H01077, 21K03583 and 23K03403.

\bibliographystyle{utphys}
\bibliography{bibnewhelas}

\end{document}